\documentclass[submission,copyright,creativecommons]{eptcs}
\providecommand{\event}{PrePost 2017} % Name of the event you are submitting to
\usepackage{breakurl}             % Not needed if you use pdflatex only.
\usepackage{listings}
\usepackage{graphicx}
\usepackage{times}
\usepackage{microtype}
\usepackage[bbgreekl]{mathbbol}

\usepackage{pgfplots}
\usepackage[textsize=footnotesize,shadow]{todonotes}

\usepackage{stmaryrd}
\usepackage[floatrow]{chemstyle}
\usepackage{graphics}
\usepackage[font=small]{caption}
\usepackage{subcaption}
\usepackage{amsfonts}
\usepackage{amssymb}
\usepackage{xcolor}
\usepackage{color}
\usepackage{graphicx} 
\usepackage{tikz}
\usepackage{mathrsfs}
\usepackage{pdflscape}
\usepackage{verbatim}
\usepackage[fleqn]{mathtools}
\usepackage{array}
\usepackage{caption}
\usepackage[inference]{semantic}
\usepackage{proof}
\usepackage{placeins}
\usepackage{enumitem}
\usepackage{bbold}
\usepackage{centernot}
\usepackage{mathtools}
\usepackage{listings}
\usepackage{algorithm}
\usepackage{algorithmicx}
\usepackage[noend]{algpseudocode}
\usepackage{tikz-qtree}
\usepackage{tikz-qtree-compat}
\usepackage{enumitem}
\usepackage{hyperref}
\usepackage{algorithmicx}
\usepackage{mathtools}
\usepackage{mathpartir}

\usetikzlibrary{automata, positioning,calc,decorations.pathmorphing,shapes,patterns,decorations.markings,arrows,fit,positioning}
\tikzstyle{vertex}=[draw,circle,minimum size=20pt]

\usepackage{times}
\usepackage{tikz}

\usetikzlibrary{automata, positioning,calc, decorations.text,shapes, fit}

\newcommand{\SSigma}{\mathbb{\Sigma}}
\newcommand{\bool}{\mathbb{B}}

\newcommand\p{\textit{p}}

\newcommand\e{\textit{e}}
\newcommand\es{\textit{es}}
\newcommand\q{\textit{q}}
\newcommand\Q{\textit{Q}}
\newcommand\B{\textit{B}}

\newcommand\rt{\textit{rt}}
\newcommand\st{\textit{st}}

\newcommand\ST{\textit{ST}}
\newcommand\Obj{\textit{Obj}}

\newcommand\OId{\textit{ObjId}}
\newcommand\oid{\textit{objId}}
\newcommand\D{\textit{D}}

\newcommand\concretedelta{\delta}
\newcommand\approximatedelta{\Delta}

\newcommand\mydef{\mathrel{\overset{\makebox[0pt]{\mbox{\normalfont\tiny\sffamily def}}}{=}}}

\newcommand\may{{\equiv_{\mbox{\normalfont\tiny\sffamily may}}}}
\newcommand\notmay{{\ \not\equiv_{\mbox{\normalfont\tiny\sffamily may}}\ }}

\newcommand{\reachable}[1]{\mathcal{R}(#1)}

\usepackage{amsthm}
\theoremstyle{definition}
\newtheorem{definition}{Definition}[section]
\newtheorem{theorem}{Theorem}[section]
\newtheorem{proposition}{Proposition}[section]

\newcommand*\xoverline[2][0.75]{%
    \sbox{\myboxA}{$\m@th#2$}%
    \setbox\myboxB\null% Phantom box
    \ht\myboxB=\ht\myboxA%
    \dp\myboxB=\dp\myboxA%
    \wd\myboxB=#1\wd\myboxA% Scale phantom
    \sbox\myboxB{$\m@th\overline{\copy\myboxB}$}%  Overlined phantom
    \setlength\mylenA{\the\wd\myboxA}%   calc width diff
    \addtolength\mylenA{-\the\wd\myboxB}%
    \ifdim\wd\myboxB<\wd\myboxA%
       \rlap{\hskip 0.5\mylenA\usebox\myboxB}{\usebox\myboxA}%
    \else
        \hskip -0.5\mylenA\rlap{\usebox\myboxA}{\hskip 0.5\mylenA\usebox\myboxB}%
    \fi}
\makeatother

\usepackage{amsthm}

\newcommand\doubleplus{+\kern-1.3ex+\kern0.8ex}

\author{
Shaun Azzopardi \qquad\qquad Christian Colombo
\email{\quad shaun.azzopardi@um.edu.mt \quad\qquad christian.colombo@um.edu.mt}
\\ Gordon J. Pace
\email{\qquad gordon.pace@um.edu.mt}
\institute{Department of Computer Science, University of Malta, Msida, Malta}
}
\title{Control-Flow Residual Analysis for Symbolic Automata\thanks{This research has received funding from the European Union's Horizon 2020 research and innovation programme under grant number 666363.}
}

\def\titlerunning{Control-Flow Residual Analysis for Symbolic Automata}
\def\authorrunning{S. Azzopardi, C. Colombo \& G.J. Pace}

\begin{document}
\maketitle

\begin{abstract}
Where full static analysis of systems fails to scale up due to system size, dynamic monitoring has been increasingly used to ensure system correctness. The downside is, however, runtime overheads which are induced by the additional monitoring code instrumented. To address this issue, various approaches have been proposed in the literature to use static analysis in order to reduce monitoring overhead. In this paper we generalise existing work which uses control-flow static analysis to optimise properties specified as automata, and prove how similar analysis can be applied to more expressive symbolic automata - enabling reduction of monitoring instrumentation in the system, and also monitoring logic. We also present empirical evidence of the effectiveness of this approach through an analysis of the effect of monitoring overheads in a financial transaction system.
\end{abstract}

\section{Introduction}
%1. (Intro) Verification of programs is becoming even more important, especially if software engineers want to live up to their name of engineers. This is especially important for sensitive programs. Hardware -> model checking already; software left behind. Static analysis of a program not always possible, since some values only available at runtime (e.g. checking that no transaction happens above a certain amount). Runtime verification however adds overheads to a program. We can do some static analysis to prove parts of a property for the whole program, or for a part of it. This can lead to a reduction of overheads at runtime.

The need for verification of a system to be able to make some guarantees about execution paths, and going beyond sampling of such paths (as done in testing), is required for critical or sensitive software (e.g.\ financial software \cite{opesefm}). The literature can be largely split into two main approaches: (i) full \emph{a priori} verification of all possible execution paths through model checking, static analysis and similar techniques, and (ii) \emph{on-the-fly} verification of execution paths to ensure that any potential violation can be immediately truncated as in runtime verification. %\gp{We need to explain what comes from Bodden's work and what is ours.}

%A solution development lifecycle usually starts from specification of how a solution should behave to an implementation and testing of this with respect to the specification. Testing of certain cases is used to identify bugs within the implementation (i.e. behaviour not respecting the specification), and through several iterations, provide an adequate solution. However such a solution is adequate only with respect to the test cases, and does not come with any dependable guarantees outside of these. This is perhaps enough for certain industries (e.g. gaming), however other industries require a better guarantee to adopt sensitive software (e.g. financial software \cite{opesefm}) that is not guaranteed to be well-behaved. Formal verification of software can aid to provide such a guarantee.

The former approaches tend not to scale to more complex and large software, which is typically addressed by abstraction techniques, e.g.\ verifying the property against an over-approximation of the program (if the over-approximation cannot violate the property, then the program cannot either), due to which the analysis is no longer complete. These approaches readily scale up to handle larger systems, and have the additional advantage that they can also deal with constraints on the environment which can only be verified fully at runtime (e.g.\ two methods of an API are never called in sequence by an unknown client application). The downside is, however, that the additional checks introduced typically add significant runtime overheads \cite{dwyer2}. Traditionally, these two approaches have been seen as alternatives to each other, although their possible complementarity has started to be explored in recent years \cite{staticrv2,bodden,dwyer}.%\gp{Shaun, add some references, here}

Approaches exist that use static analysis to prove parts of a property with respect to a program, such that either the whole property is proved statically or it is pruned such that there is less to monitor at runtime, or vice-versa so that certain parts of the program are proved safe to not monitor. We call this \emph{residual analysis}, with the pruned property called a \emph{residual}. In language-theoretic terms, the language of a residual intersected with that of a program is equal to the language of the original property intersected with the program. This paper deals with the creation of such residuals in the presence of properties parametrised over different objects with different behaviour. 

In particular, Clara \cite{bodden} is one such approach, acting on purely control-flow properties (without any consideration for data state) defined as automata with transitions triggered by method calls in Java programs. Through an analysis of the source code, Clara can be used to determine whether a property transition can never be taken by the program and whether parts of a program can be safely unmonitored. 
%CC suppressing to gain space 13/4/17
%Removing such transitions from a property creates a residual property that is smaller than, or equivalent to, the original property and since some parts of the program will not activate the monitor any longer, then there will be less overheads at runtime due to monitoring. 
The approach uses three analysis steps, each equivalent to a comparison of the property automaton with a finer over-approximation of the control-flow of a program. However, in practice, one frequently desires properties which are more expressive than these simple automata (e.g. one may want to talk about the value of a transaction after some sequence of events). 

DATEs (Dynamic Automata with Events and Timers) 
\cite{Colombo2009} involve symbolic automata possibly running in parallel, with transitions triggered by either a method invocation or a timer event, and conditioned on a boolean expression on variables specific to the monitor and the program. Moreover, when triggered, a transition can also perform an action which may also affect the triggering of other transitions (e.g.\ increase some internal counter used by another transition's condition). In this paper, we do not handle timers and the dynamic creation of DATEs, and focus on extending Clara to automata with transitions guarded by conditions and actions as an initial step towards handling full DATEs. Even with this limitation, applying Clara directly to DATEs proves to be unsound, as we shall discuss informally, given transitions with side-effects may be removed (possibly changing the verdict of the reduced property).  
%CC suppressing to gain space 13/4/17
%DATEs are used within the LARVA (Logical Automata for Runtime Verification and Analysis) tool to specify the security properties for a Java system. 

Our contribution in this paper is two-fold: (i) extending the intuition behind Clara's first two analyses to produce both a residual DATE and a residual instrumentation of the program, and (ii) a novel analysis that uses a control-flow graph of a program to determine if any transitions in a DATE can (or can not) be reached by the program. In a case study, we show that each of these analyses can produce significant reduction in runtime overheads (down to 4\% from an average of 97\% of the original unmonitored run time), we also explain in which cases such results can be expected. Full proofs of the results presented in this paper can be found in \cite{technicalreport}.%We have incorporated these analyses in a tool which we call Clarva.

In presenting our results, we first detail some formal preliminaries, using simple automata with events in Section \ref{s:background} and use these to discuss Clara's analyses, while in Section \ref{s:date-analysis} we present DATEs and define residuals over them, which we evaluate in Section \ref{s:case-study}. We discuss related work in Section \ref{s:related-work}, and conclude proposing future work in Section \ref{s:conclusions}. Due to space restrictions, the proofs of the results presented in this paper are not included. However, they are available in a technical report available online \cite{technicalreport}.

%In Section \ref{s:background} we briefly introduce some formal notation and informally discuss Clara, while we present our generalisation of it for DATEs in Section \ref{s:date-analysis}. In Section \ref{s:case-study}, a case study involving a financial transaction system is considered along with some properties as a backdrop against which we discuss the utility of the presented analysis. We present related work in Section \ref{s:related-work}, and conclude proposing future work in Section \ref{s:conclusions}.
\section{Programs' Runtime Traces and Abstractions}
\label{s:background}
%2. (Background) Properties -> automata on events, logic (JML). Bodden did this for Clara: explain informally and using diagrams and examples. Explain other forms of static analysis, and the summarising program residual. However properties can be more complex, e.g.\ DATEs and ppDATEs; flow into StaRVOOrS. (Quote ISoLa paper on residuals and SEFM paper). Soot, callgraphs and control-flow graphs.
%Properties can be typically classified as \emph{control-oriented}, i.e.\ pertaining to the path through the program taken by the execution e.g.\ specifying that some methods should not be called in some sequence; or \emph{data-oriented} i.e.\ placing constraints on the values of the data flow in the system e.g.\ specifying pre- and post-conditions on some variable value before and after a method call; or even possibly combining both \cite{staticrv1}. 

%\subsection{Properties, Traces and Programs}
In this section we look at safety properties over the control-flow of a program, written in the form of automata with transitions being triggered by the program\footnote{It is worth noting, that although we will allow for branching on data values in our formalism, we do not trigger transitions over changes in data values, hence our characterisation of these properties as being over the control-flow of the program.}. We will build on this formalism in the rest of the paper. 

%In this section we look at control-flow safety properties written in the form of automata, with transitions being triggered by the program (e.g. by a method invocation), we call these program actions \emph{event generators}, since they generate monitor events. We will build on this formalism in the rest of the paper.%objects => replication => overheads

\begin{definition}[Property automata]
A \emph{property automaton} $\pi$ is a tuple $\langle \Q, \Sigma, \q_0, \B, \delta \rangle$, where $\Q$ is a finite set of states, $\Sigma$ is a set of events, $\q_0$ is the initial state ($\q_0\in Q$), $\B$ is the set of bad states ($\B\subseteq Q$), and $\delta$ is the transition relation ($\delta \subseteq \Q\times \Sigma\times \Q$), which is deterministic and total with respect to $\Q\times \Sigma$. 

We write $\q \xrightarrow{\e}_\pi \q'$ for $(\q,\e,\q') \in \delta$, $\q \xRightarrow{\es}_\pi \q'$ for the transitive closure of $\delta$ (with $\es \in \Sigma^*$), and $\q \hookrightarrow_\pi \q'$ to denote that $\q'$ is reachable from $\q$ (i.e.\ $\exists \es\cdot q_0 \xRightarrow{\es}_\pi \q'$). We leave out $\pi$ if it is clear from the context.%, and similarly that a transition $t'$ is reachable from $t$ by $t \hookrightarrow t'$. .

Finally, we will write $\pi \upharpoonright \Sigma'$ to denote the property automaton identical to $\pi$ except that the alphabet is restricted to $\Sigma'$: $\Sigma_{\pi \upharpoonright \Sigma'} \mydef \Sigma'$ and $\delta_{\pi \upharpoonright \Sigma'} \mydef \{(q,e,q') \in \delta_\pi \mid e \in \Sigma'\}$); and $\pi \upharpoonright \delta'$ as $\pi$ with the transition relation restricted to $\delta'$: $\delta_{\pi \upharpoonright \delta'} \mydef \delta_\pi \cap \delta'$.
\end{definition}

Consider that in monitoring we are concerned with points of interest during the execution of the program (e.g. when a method is called), which correspond to particular statements or regions in a program's source code. We use these corresponding program statements to trigger events at runtime to enable monitoring --- and calling them \emph{event generators}, since they generate property events. The problem of verification is then, given a property automaton and a program, to ensure that all program traces generated do not transition into a bad property state \cite{leucker}. We call $t \in \Sigma^*$ a \textit{ground trace}, %\cc{while call them so?}\sa{because Bodden calls them so} 
while we denote the set of events appearing in $t$ by $\Sigma(t) \mydef \{e \mid \exists i \in \mathbb{N} \cdot t(i) = e\}$, overloaded to sets of ground traces $T \subseteq \Sigma^*$. 
%CC suppressing to gain space 13/4/17
%A ground trace then satisfies a given automaton property if it does not go through a bad state.

\begin{definition}[Property satisfaction]
A ground trace $t \in \Sigma^*$ is said to satisfy property automaton $\pi$ if no prefix of $t$ leads to a bad state from the initial state:
$t \vdash \pi \mydef \forall t' \in \textit{prefixes}(t) \cdot \nexists \q_\B \in \B \cdot \q_0 \xRightarrow{t'} \q_\B
$. We overload this notation to sets of ground traces $T \vdash \pi$ to indicate that all traces in $T$ satisfy $\pi$.
\end{definition}

\begin{figure}[ptb]
\scalebox{0.7}{
  \begin{tikzpicture}[every text node part/.style={align=center},node distance=6cm,on grid,auto]
	%States
	\node (s_0) [initial,
	% initial text={\textbf{For each:} Stream s},
	 circle, draw] [] {$\q_a$};
	\node (s_1) [circle, draw] [right = 3cm of s_0]{$\q_b$};
	\node (s_2) [accepting, circle, draw, fill=red] [below = 2cm of s_0]{$\q_c$};
	\node (s_11) [circle, draw] [right = 3cm of s_1]{$\q_d$};
	\node (s_22) [accepting, circle, draw, fill=red] [below = 2cm of s_11]{$\q_e$};

	%Transitions
	\path[->] [loop above] (s_0)	edge node {*} (s_0); %{\textit{close(),}\\ \textit{read()}} (s_0);
	\path[->] [bend right=30] (s_0)	edge node {\textit{open()}} (s_1);
	\path[->] [bend right=30] (s_1)	edge node {\textit{close()}} (s_0);
	\path[->] [right] (s_0)	edge node {\textit{write()}} (s_2);
	\path[->] [right] (s_2)	edge node {\\\textit{open()}} (s_1);
%	\path[->] [bend left=30] (s_2)	edge node {\textit{close()}} (s_0);
	\path[->] [loop left] (s_2)	edge node {*} (s_2); %{\textit{write(),}\\ \textit{read()}} (s_2);
	\path[->] [loop above] (s_1)	edge node {*} (s_1);%{\textit{write()}} (s_1);
	\path[->] [bend right=30] (s_1)	edge node {\textit{read()}} (s_11);
	\path[->] [bend right=30] (s_11)	edge node {\textit{read()}} (s_1);
	\path[->] [left] (s_11)	edge node {*} (s_22); %{\textit{s.write(),}\\ \textit{s.close()}} (s_22);
	\path[->] [loop above] (s_11)	edge node {\textit{lookAhead(),}\\ \textit{open()}} (s_11);
	\path[->] [loop right] (s_22)	edge node {*} (s_22);
	\end{tikzpicture}  
}
	\captionof{figure}{Property disallowing writing on a closed stream, and writing or closing while in the middle of an odd number of reads.}
  \label{fig:typestatepropwithread}
\end{figure}
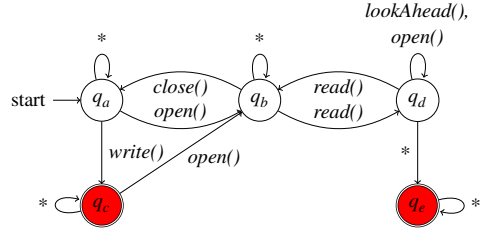

Consider as an example the property automaton shown in \ref{fig:typestatepropwithread}, which specifies that the \textit{write} method cannot be called before \textit{open} is called, and that the \textit{read} method is called in pairs\footnote{Note that in this example there are outgoing transitions from bad states, although any trace that goes through a bad state is judged as violating. Clara's and our semantics allow this, since we may want to count the number of violations, or in the case of DATEs transitions may actually repair the violation (although we would still want to note it).}. Bad states are marked in red, while an asterisk (*) on a transition is syntactic sugar used to denote that if at that state an event happens for which no other transition matches, then the asterisk transition is taken.%In the rest of the paper we may omit such transitions looping at the same state, but assume that if no transition exists at a state for a certain event then it has no effect (i.e. we transition into the same state).

%Property automata can be restricted to a smaller alphabet or transition relation by dropping a number of transitions.%\cc{felt i had to add this last part}

%\begin{definition}[Property restriction]
%The \emph{alphabet-reduction} of $\pi=\langle\Q, \Sigma, \q_0, \B, \delta\rangle$ to alphabet $\Sigma'$ (with $\Sigma' \subseteq \Sigma$), written $\pi\upharpoonright_a \Sigma'$ is defined to be $\langle\Q, \Sigma, \q_0, \B, \delta\upharpoonright_a \Sigma'\rangle$, where $\delta\upharpoonright_a \Sigma'$ is defined to be: $\delta \cap (Q\times \Sigma'\times Q)$. %\gp{Perhaps we should use the work \emph{restriction} instead of \emph{reduction}.}
%
%Similarly, the \emph{transition-reduction} of $\pi$ to transition relation $\delta'$, written $\pi\upharpoonright_t \delta'$ is defined to be $\langle\Q, \Sigma, \q_0, \B, \delta\cap \delta'\rangle$. We will leave out the $a$ and $t$ when they are clear from the context. %\gp{I've renamed $\upharpoonright$ and $\upharpoonright$ to $\upharpoonright_a$ and $\upharpoonright_t$, but we should probably overload both to the single symbol $\upharpoonright$. Next time, define the symbol, Shaun, so that we can change symbols easily.}
%\end{definition}

In practice, one would want to instantiate a property automaton for every instance of the object being verified. For example, the property automaton shown in \ref{fig:typestatepropwithread} should ideally be monitoring for every stream in use. To enable such replication of property automata, we extend them to parametric properties \cite{parametricprops} (or typestate automata \cite{bodden}). To generalise property automata to handle parametrisation, we start by extending traces to parametrised traces in which each event is associated with an identifier of the object\footnote{Although we will be using the term `object' in this paper, events can be parametrised with respect to an identity other than the object on which the method is invoked.} to which the event pertains. 
Note that we allow an alphabet to be parametrised by a set of identifiers $\alpha$ with: $\SSigma_\alpha \mydef \alpha \times \Sigma$. Traces over such alphabets will be referred to as parametrised traces.%Such parametrised property automata are effectively typestate automata \cite{bodden}. At runtime it is also possible to consider when two identifiers refer to the same object, which we simulate through an equivalence relation between identifiers.
%Thus, at runtime, if \textit{open} and \textit{write} are called in succession we cannot immediately conclude non-violation of the property, since they may have been called on different objects (Java objects or with respect to some data).

A trace contains events possibly related to different objects. To consider satisfaction of a property by one object we project the trace onto that object (assuming an equivalence relation between the objects).
% can be lifted over parametrised traces by checking for the satisfaction of each possible projection.

%\begin{definition}
%A ground trace can be reduced with respect to an alphabet $\Sigma$ by removing any event not in $\Sigma$.
%\begin{align*}
%\langle \rangle \upharpoonright \Sigma &\mydef \langle \rangle\\
%\e\colon\textit{es}\upharpoonright \Sigma &\mydef \e\colon (\textit{es} \upharpoonright \Sigma) &&(\e \in \Sigma)\\
%\e\colon\textit{es}\upharpoonright \Sigma &\mydef (\textit{es} \upharpoonright \Sigma) &&(\e \not\in \Sigma)
%\end{align*}
%We extend to sets of ground traces $T \subseteq \Sigma^*$.
%\begin{align*}
%T \upharpoonright \Sigma &\mydef \{t \in T \bullet t \upharpoonright \Sigma\}
%\end{align*}
%\end{definition}

%A runtime trace may then have an interleaving of events parametrised by different objects, which can be projected (or sliced) into a ground trace for each object.

\begin{definition}[Parametrised traces]
%A runtime trace parametrised over a type of identifiers $\alpha$ is a finite sequence of parametrised events: $\rt \in \SSigma_\alpha^*$.
The projection of a parametrised trace $\rt \in \SSigma_\alpha^*$ with respect to an identifier $x \in \alpha$ and an equivalence relation between objects ${}\equiv{} \in \alpha \leftrightarrow \alpha$, written $\rt \downarrow x$, is defined to be the sequence of items in $\rt$ with identifiers equivalent to $x$, as follows 
\footnote{We use the standard notation $x:xs$ to denote the list with head $x$ and tail $xs$.}:
\small\[\begin{array}{lcl}
\langle \rangle \downarrow x &\mydef & \langle \rangle\\
((x',\p)\colon \rt) \downarrow x &\mydef&
\left\{\begin{array}{ll}
(x',\p)\colon (rt \downarrow x) & \mbox{ if $x\equiv x'$}\\
rt \downarrow x & \mbox{ otherwise}
\end{array}\right.
\end{array}\]
\normalsize

A parametrised trace $\rt \in \SSigma^*$ is said to satisfy a property automaton $\pi$, written $\rt \Vdash \pi$, if for each identifier 
%\cc{better say identifier instead of object no?} 
$x\in \alpha$, the projection of $\rt$ onto $x$ satisfies the property: $\rt \Vdash \pi \mydef \forall x\in \alpha \cdot \rt \downarrow x \vdash \pi$.
%We overload this notation over sets of parametrised traces:.% $\RT \subseteq \SSigma_\alpha^*$.

%The notation is overloaded for sets of parametrised runtime traces.%: if  $\RT \subseteq \SSigma_\alpha^*$, we write $\RT\downarrow_\equiv x$ to denote the projection of each trace in $\RT$ to identifier $x$.
%\cc{why mention $\ob$?}
%We will simply write $\downarrow$ when $\equiv$ is clear from the context.
%\begin{align*}
%\RT \downarrow \ob &\mydef \{\rt \in \RT \bullet \rt \downarrow \ob\}
%\end{align*}
\end{definition}

At runtime, we have perfect knowledge of the equivalence relation between paramet\-rised events. However, when using static analysis, this is not always possible. %For instance, %control flow analysis typically takes the source code structure into consideration, and thus, for example, a particular method invocation in a loop might be related to different objects in different iterations of the loop. Many static analysis techniques on the system code allow us to generate traces with partial information as to which method calls may (or may not) refer to the same object. 
In the case of parametrisation by objects, several variable alias analyses exist, that can give partial information on whether two source code variables can point to the same runtime object (we use \cite{aliasanalysis}). In such cases, we have three possible outcomes: (i) the two variables always refer to the same object; (ii) the two variables always refer to different objects; and (iii) neither of the previous two cases can be concluded. In the literature, this information is typically encapsulated in two relations --- a \emph{must} relation $\equiv$, which relates two event generators (e.g. method calls) if their objects always (must) match, and a \emph{may} relation $\may$, which relates two event generators if their objects may match, with the former relation being a subset of the latter \footnote{Other analyses would be applicable in the case of parametrization by data instead of objects, and when this is not possible all identifiers can be related by the may relation soundly}. Thus, with such variables as static object identifiers, when statically we are given a trace of event-identifier couples, we can extract the possible runtime traces generated by these, through projection on certain identifiers. This is in contrast to parametrised traces where the behaviour of each object is perfectly known and thus only one ground trace is generated through projection.

\begin{definition}[Static Parametrised Traces]
A static parametrised trace is a parametrised trace $\st\in\SSigma_\alpha^*$, with two relations over $\alpha$: (i) a must-alias equivalence relation $\equiv{} \in \alpha \leftrightarrow \alpha$, and (ii) a may-alias relation $\may \in \alpha \leftrightarrow \alpha$, such that $\equiv{} \subseteq \may$. We define the projection
%\footnote{This corresponds to the notion of program slicing, by slicing the instrumented program statements according to whether they can be associated with the same object or not.}%
 of a static parametrised trace $\st$ with respect to parameter $x$, written $\st \Downarrow x$, as follows:
\small\[\begin{array}{lcl}
\langle \rangle \Downarrow x &\mydef& \{\langle \rangle\}\\
((x', \e)\colon \st) \Downarrow x & \mydef &
\left\{\begin{array}{ll}
  \{\e\colon\es \mid \es \in \st \Downarrow x\}       & \mbox{ if $x\equiv x'$}         \\
  \st \Downarrow x               & \mbox{ if $x \notmay x'$}\\
  \st \Downarrow x \cup \{\e\colon\es \mid \es \in \st \Downarrow x\} & \mbox{ otherwise}
\end{array}\right.
\end{array}\]
\normalsize
%\[\begin{array}{lcl}
%(\langle \rangle, \may) \Downarrow \oid &\mydef& \{\langle \rangle\}\\
%((\textit{ip}, \oid', \e)\colon \ps, \may) \Downarrow \oid & \mydef &
%\left\{\begin{array}{ll}
%  \{\e\colon\es' \mid \es' \in \ps \Downarrow \oid\}       & \mbox{ if $\oid=\oid'$}         \\
%  \ps \Downarrow \oid \cup \{\e\colon\es' \mid \es' \in \ps \Downarrow \oid\} & \mbox{ if $\oid \may \oid'$} \\
%  \ps \Downarrow \oid               & \mbox{ otherwise}
%\end{array}\right.
%\end{array}\]
%\begin{align*}
%(\langle \rangle, \may) \Downarrow \oid &\mydef \{\langle \rangle\}\\
%((\textit{ip}, \oid, \e)\colon \st, \may) \Downarrow \oid &\mydef \{\e\colon\st' \mid \st' \in \st \Downarrow \oid\}\\
%((\textit{ip}, \oid', \e)\colon \st, \may) \Downarrow \oid &\mydef \{\st', \e\colon\st' \mid \st' \in \st \Downarrow \oid\} &&(\oid \may \oid') \\
%((\textit{ip}, \oid', \e)\colon \st, \may) \Downarrow \oid &\mydef \{\st' \mid \st' \in \st \Downarrow \oid\} &&(\oid \maynot \oid')
%\end{align*}

We overload this notation over sets of static parametrised traces: $\ST \Downarrow x \mydef \bigcup_{\st \in \ST} \st \Downarrow x$; and over sets of identifiers: $\ST \Downarrow X \mydef \bigcup_{x \in X} \ST \Downarrow x$.

Trace $\st$ is said to satisfy property automaton $\pi$ , written $\st \Vvdash \pi$, if for each identifier $x\in \alpha$ the projection of $\st$ onto $x$ satisfies the property: $\st \Vvdash \pi \mydef \forall x \in \alpha \cdot \st \Downarrow x \vdash \pi$. 
%
%
%xxx
%We overload this notation to work over sets of static parametrised traces: if $\ST \subseteq \SSigma_\alpha^*$, we write $\ST \Downarrow x$ to denote the projection of each trace in $\ST$ to identifier $x$. We similarly overload the notation to write $\st \Downarrow X$ and $\ST \Downarrow X$ to denote the projection to events pertaining to any identifier in set $X$.
%We extend this for a set of traces $\ST \subseteq \SSigma^*$, both over one or a set of object identifiers.
%\begin{align*}
%(\ST, \may) \Downarrow \oid &\mydef \bigcup_{\st \in \ST} (\st, \may) \Downarrow \oid\\
%(\ST, \may) \Downarrow \OId &\mydef \bigcup_{\oid \in \OId} (\ST, \may) \Downarrow \oid
%\end{align*}
\end{definition}
It is worth noting that although we are only considering parametrisation over a single identifier (e.g. to an object), a property can be parametrized over multiple objects. The work presented here applies to that case, by abstracting tuples of identifiers into a single identifier, with point-wise must and may-alias relations.

%It will be important for us to consider the set of events used by a trace.
%
%\begin{definition}
%The alphabet of a static parametrised trace is the set of events it uses.
%\begin{align*}
%\Sigma(\langle \rangle) &\mydef \{\}\\
%\Sigma((\textit{ip}, \oid,\e) \colon \st) &\mydef \{\e\} \cup \Sigma(\st)
%\end{align*}
%We also overload this to sets of traces, $\ST : 2^{(\IP \times \OId \times \Sigma)^*}$.
%\begin{align*}
%\Sigma(\ST) \mydef \bigcup_{\st \in \ST} \Sigma(\st)
%\end{align*}
%\end{definition}

%\subsection{Programs}
\noindent We now turn our view from individual traces to the programs which generate them.

%Using this definition of traces we can define what we mean by a program at runtime and a static approximation of it. Note that for static approximations of program we also introduce a function that associates an instruction pointer ($\textit{ip} \in \IP$) with each object identifier, this identifies the instruction in the source code that activates the events associated with the identifier. This will be used later to enable several instructions to be silenced such that they no longer activate any event.

\begin{definition}[Programs]
For a program $P$ over an alphabet $\Sigma$ with a set of runtime objects $\Obj$ and static object identifiers $\OId$, (i) we will write $P_\Sigma^R$ to denote the set of parametrised traces over $\Obj$ i.e.\ $P_\Sigma^R\subseteq \SSigma_\Obj^*$ with equivalence $\equiv$ over $\Obj$; (ii) we will write $P_\Sigma^S$ to denote the set of static parametrised traces over $\OId$ i.e.\ $P_\Sigma^S\subseteq \SSigma_\OId^*$ with relations $\equiv$ and $\may$.

%
% the concrete runtime traces, written $P_\Sigma^R$ to denote the set of traces produced by the program at runtime parametrised over object pointers in $\Obj$ over which equivalence relation $\equiv$ is defined: $P_\Sigma^R \subseteq \SSigma_\Obj^*$; (b) the static approximation of a program, written $P_\Sigma^S$ to denote the set of static traces parametrised over object identifiers $\OId$ \cc{can we find something which doesn't look like `Old'? :)}over which must-alias  and may-alias relations $\equiv$ and $\may$ are defined: $P_\Sigma^S \subseteq \SSigma_\OId^*$.
%We overload the projection and satisfaction operators to such programs, using the required set of traces and the needed relations.
\end{definition}

%\cc{did some changes in this area.. please check Shaun}
In what follows, we assume that the static parametrised trace generator is an over-approximation of the parametrised trace generator: $P_\Sigma^R \downarrow \Obj \subseteq P_\Sigma^S \Downarrow \OId$.

In our effort to reduce overheads by reducing a property, we will need to ensure that monitoring the program by the created residual is enough, i.e. both the original property and the residual give the same verdict for any given program trace, although their results may vary for other traces.
%The approaches we will explore in this paper involve property-specific transformations of program monitoring in order to reduce overheads in such a manner that will not affect the verdict with respect to that particular property. Next, we define notions of trace and program equivalence with respect to a property or program.

%
%\begin{definition}
%Two static approximations of a program, $P^{S}_{\Sigma}$ and $P^{S'}_{\Sigma'}$, are said to be equivalent with respect to a property $\pi$ (with $\Sigma, \Sigma' \subseteq \Sigma_\pi$) if the satisfaction of one's behaviour implies the other's.
%\begin{align*}
%P^S_\Sigma \cong_\pi P^{S'}_{\Sigma'} \mydef P^S_\Sigma \Vvdash \pi \iff P^{S'}_{\Sigma'} \Vvdash \pi
%\end{align*}
%\end{definition}

\begin{definition}[Equivalence]
\label{equivwrttraces}
Two properties $\pi$ and $\pi'$ are said to be equivalent with respect to a set of ground traces $T$, written $\pi \cong_T \pi'$, if every trace in $T$ is judged the same by either property:
%\begin{align*}
%\pi \cong_{T} \pi' \mydef T \vdash \pi \iff T \vdash \pi'
$\forall t \in T \cdot t \vdash \pi \iff t \vdash \pi'$.
%\end{align*}
%Similarly, for a statically approximated program, $P^S_\Sigma = (\ST, \equiv, \may, \mathcal{I})$.
%\begin{align*}
%\pi \cong_{P^S_\Sigma} \pi' \mydef \forall \st \in \ST \cdot \st \Vvdash \pi \iff \st \Vvdash \pi'
%\end{align*}
This is lifted to parametrised and static parametrised traces, and sets thereof.
%This notion is lifted to sets of abstract traces and thus static approximations of program with
%$\pi \cong_{P^S_\Sigma} \pi'$ being defined to be $\forall \st \in \ST \cdot \st \Vvdash \pi \iff \st \Vvdash \pi'$.
\end{definition}

%\begin{proposition}
%\label{basic-equivalence-properties}
%Equivalence with respect to a set of ground traces is (i) an equivalence relation; and (ii) preserved with respect to reduction of the set of traces: if $\pi \cong_{T'} \pi$ and $T \subseteq T'$, then $\pi \cong_T \pi$. 
%\end{proposition}

It is straightforward to prove that equivalence up to traces is an equivalence relation and is preserved when reducing the set of traces, from which the next property holds.

\begin{proposition}
\label{equivintersection}
If $\pi \cong_T \pi'$ and $\pi' \cong_{T'} \pi''$, then $\pi \cong_{T \cap T'} \pi''$.
\end{proposition}

We can also show that equivalence with respect to the static parametrised trace generator can be expressed in terms of its projection onto ground traces. %, following from Prop. \ref{staticprogesatequivtoprojectionsat1}, which allows us to consider just equivalence between such traces in the rest of the paper.

\begin{proposition}
\label{staticprogesatequivtoprojectionsat2}
%Two properties are equivalent with respect to a statically approximated program if and only if the properties are equivalent with respect to the projection of the approximation onto the set of object identifiers:
%\begin{align*}
$\pi \cong_{P^S_\Sigma} \pi' \iff \pi \cong_{P^S_\Sigma \Downarrow \OId} \pi'$.
%\end{align*}
\end{proposition}

As we have discussed, some parts of the program may be proved safe to unmonitor, upon which we can silence these parts and transform the static parametrised trace generator by turning off certain identifier-event pairs.

\begin{definition}
Given a static parametrised trace $\st$, and a set of identifier-event pairs $E \in 2^{\OId \times \Sigma}$, the silencing of $\st$ by $E$, written, $\textit{silence}(\st, \textit{ips})$, is defined to be the original trace $\st$ except for elements in $E$:
\[\begin{array}{lcl}
\textit{silence}(\langle \rangle,E) &\mydef &\langle \rangle\\
\textit{silence}((x,\e)\colon \st,E) & \mydef &
\left\{\begin{array}{ll}
  \textit{silence}(\st,E)       & \mbox{ if $(x,\e) \in E$}         \\
  (x,\e)\colon \textit{silence}(\st,E)  & \mbox{ otherwise}\\
\end{array}\right.
\end{array}\]
%We overload this notation to sets of static parametrised traces.
\end{definition}

We can also define equivalence between a program and its transformation (see \cite{technicalreport}), however given lack of space we focus solely on equivalence between residuals and the original property in this paper, but define silencing since Clara
%\begin{proof}
%This follows from Prop. \ref{staticprogesatequivtoprojectionsat1} and Defn. \ref{staticsatdef}.
%\end{proof}

%In the next sub-section we shall now look at Clara's algorithms informally, in the context of the theory just presented.

%\subsection{Clara}

%\sa{Mention how Clara (and we) use Soot to analyse programs}

%Clara analyses properties with respect to programs to potentially turn off some method invocations in the program, such that they no longer trigger any monitor at runtime.

%\begin{definition}
%A program $P$, with respect to a property $\pi$, taking into account statically approximated objects, can be silenced with respect to a set of events by removing events not in that set from every trace:
%
%\begin{gather*}
%\textit{silence}(t, \Sigma) \mydef \left\{
%	\begin{aligned}
%		&\textit{head}(t);\textit{silence}(tail(t), \Sigma)   && \mbox{if } \textit{head}(t) \in \Sigma \\
%		&\textit{silence}(tail(t), \Sigma) && \mbox{otherwise}
%	\end{aligned}
%\right.\\
%\textit{silence}((ts, \textit{must}, \may), \Sigma) \mydef (\{t \in ts \cdot silence(ts, \Sigma)\}, \textit{must}, \may)
%\end{gather*}
%\end{definition}

%Recall how a property can specify the behaviour of objects of a certain type used by the program. However, not all objects may be used in the same way by the program, leading to the possibility that some method calls in a problem never affect the violation of a property. Clara thus can silence specific statements in a program, removing the possibility that they appear in a trace at runtime.

Based on the notions presented, we discuss the analysis techniques used in Clara \cite{bodden}, where each of its analyses reduces the points in a program that activate the monitor at runtime. The only inputs Clara needs is the source code of the program (which is then analysed using Soot \cite{soot}) and a property automaton. %, meaning that some methods may no longer be instrumented fully, or partially (with some method calls still enabled).
%\cc{include also points which trigger the original monitor but not the residual?}
The basic thesis of Clara is then that appropriate silencing of certain events, does not affect satisfiability of the program with respect to the property but reduces the length of the traces to be analysed: 
%\begin{claim}
Given a static parametrised program $P = P^S_\Sigma$ and property $\pi$, the reduced program approximation obtained through Clara, $P'=\textit{Clara}(P^S_\Sigma, \pi)$, is sound with respect to $\pi$: $P \cong_\pi P'$.
%\end{claim}

\noindent Clara uses three analysis techniques to reduce the program approximation \cite{bodden}:
%Similarly, the first analysis of Clara can be used to reduce a property by removing certain events from a property and doing a reachability analysis and thus producing a residual property.

	\noindent{\bf Quick Check.} Some events specified by the property may not correspond to any method invocations by the  program, e.g.\ consider that given \ref{fig:typestatepropwithread}, a program may only open streams and write to them, but never read from them. Also, some events may only appear on loops in the same state, and therefore never cause a change in state (e.g.\  \textit{lookAhead}). Clara's first analysis can be used to remove these kinds of events from the property, and the corresponding transitions. This may lead to some states becoming unreachable from the initial state, or states that cannot reach a bad state, and thus these can also be removed. If a bad state cannot be reached from an initial state, then the property is satisfied.

\noindent{\bf Orphan Shadows Analysis.} 
The first analysis ignores the fact that events are paramet\-rised. Consider a program where only \textit{open} and \textit{lookAhead} are ever called on one object, then by looking at the property we can note that this object can never violate it (by performing the first analysis on this object, instead of on the whole program), therefore both method calls can be silenced. This can then produce, for each object, a set of such instrumentation points that can be disabled without affecting the result of monitoring. %This is called \textit{Orphan Shadows Analysis}, where the first analysis is extended to consider objects (or rather their static approximation, as discuss in the next section).

%\begin{algorithmic}
%\Function{$\text{OrphanShadows}$}{$\mathcal{P}^S_\pi$, $\ob$}
%%\State $\textit{methods}(\ob) = \bigcup_{t \in \mathcal{P}^S_\pi} \{i < length(t) \bullet \e_i\}$ 
%\State compSyms(s) $= \{a \in \Sigma \mid \exists s' \in S \cdot compatible(s,s') \wedge a = label(s')\}$
%\State\Return $\{s \in S \mid label(s) \in \textit{QuickCheck}_{\mathcal{P}}(\textit{compSyms(s)})\}$
%\EndFunction
%\end{algorithmic}

\noindent{\bf Flow-Sensitive Nop-Shadows Analysis.} The first two analyses do not take into account any of the control-flow of the program, they just consider which methods are invoked or not but not in which order. This can be taken into account by considering a control-flow graph (CFG) of a program, which represents a superset of its event-triggering behaviour, and silence any statements that, if present or not, do not affect violation. %By intersecting a property with the CFG we can then determine which transitions in the property have an effect on the violation of the property. 
%However, constructing the whole-program CFG can be expensive \cite{technicalreport}. 
\begin{figure}[ptb]
\scalebox{0.7}{
  \begin{tikzpicture}[every text node part/.style={align=center},node distance=6cm,on grid,auto]
	%States
	\node (s_0) [initial, circle, draw, initial] {$\q_1$};
	\node (s_1) [circle, draw] [right = 3cm of s_0]{$\q_2$};
	\node (s_2) [circle, draw] [right = 3cm of s_1]{$\q_3$};
	\node (s_3) [circle, draw] [right = 3cm of s_2]{$\q_4$};
	\node (s_4) [accepting, circle, draw] [right = 3cm of s_3]{$\q_5$};
	%Transitions
	\path[->] [loop above] (s_0)	edge node {\textit{$s_1$.write}, \textit{$s_2$.open}} (s_0);
	\path[->] [loop above] (s_4)	edge node {\textit{$s_1$.write}, \textit{$s_2$.open}} (s_4);
	\path[->] [loop above] (s_3)	edge node {\textit{$s_1$.write}, \textit{$s_2$.open}} (s_3);
	\path[->] [bend left=10] (s_4)	edge node {$\epsilon$} (s_0);
	\path[->] [] (s_0)	edge node {\textbf{$s_3$.open}} (s_1);
%	\path[->] [bend left = 30] (s_1)	edge node {\textit{open}} (s_0);
	\path[->] [] (s_1)	edge node {\textbf{$s_4$.close}} (s_2);
	\path[->] [] (s_2)	edge node {\textit{$s_5$.open}} (s_3);
	\path[->] [] (s_3)	edge node {\textit{$s_8$.close}} (s_4);
%	\path[->] [bend left = 70, right] (s_2)	edge node {\textit{open}} (s_0);
%	\path[->] [bend left = 30] (s_2)	edge node {\textit{close}} (s_1);
%	\path[->] [bend left = 30] (s_0)	edge node {\textit{read}} (s_11);
%	\path[->] [bend left = 30] (s_11)	edge node {\textit{read}} (s_0);
%	\path[->] [] (s_11)	edge node {\textit{write}} (s_22);
	\end{tikzpicture}
}
%\end{minipage}
	\captionof{figure}{Example method CFG generalized to whole-method CFG.}
  \label{fig:nopshadowpropex}
\end{figure}
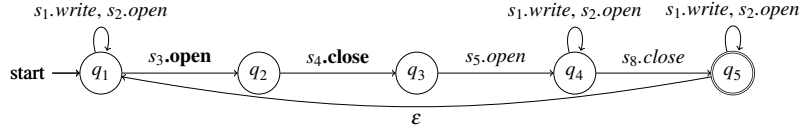

Through several over-approximations of a whole-program CFG and synchronous composition of these with the property, Bodden et\ al.\ identify sequences of instrumentation points that only ever transition from and to the same state (taking into account parametrisation of the property), with no bad states in between. Such points (which Bodden et\ al.\ call \emph{nop shadows}) never have an effect on violation, meaning they can be silenced, reducing the amount of times the monitor is triggered at runtime. 

As an example, consider the synchronous composition of the property in \ref{fig:typestatepropwithread} and the approximated CFG in \ref{fig:nopshadowpropex}, and assume that all the object identifiers ($s_i$) associated with an event always refer to the same event. 
%\cc{didn't understand what $q_a$, $q_b$, etc refer to}
One can then note that after $\q_1$, the synchronous composition is either in $\q_a$ or $\q_c$; then taking $\q_1 \xrightarrow{\textit{$s_3$.open}} \q_2$ will lead to $\q_b$, and then $\q_2 \xrightarrow{\textit{$s_4$.close}} \q_3 \xrightarrow{\textit{$s_5$.open}} \q_4$ will necessarily lead to $\q_b$.   Since, then, these two transitions do not affect the control-flow, they can be disabled such that they do not activate the monitor at runtime. Note the loops at state $\q_4$ represent the flattened behaviour of a method called at that state, while those at $\q_1$ and $\q_5$, the behaviour outside the method. %Note that if this CFG was a precise approximation of the program we could conclude that the program will violate the property at runtime (consider the looping transition at the initial state), however since it is an overapproximation detecting nop shadows is the most we can do.

\section{Control-Flow Residual Analysis of DATEs}
\label{s:date-analysis}

The properties considered by Clara are automata with explicit state. However, some properties require a richer specification language --- an extension to automata to deal with more expressive properties, 
%\sa{undefined transitions are taken to lead to an accepting state.}
 %however these can be extended to allow for more expressive properties. One such extension of automata 
 are DATEs %(Dynamic Automata with Timed Events)
  \cite{Colombo2009}. One way in which DATEs extend finite state automata is through the introduction of a symbolic state which can be checked and updated on transitions which trigger on \textit{events}, with \emph{conditional guards,} and perform \textit{side-effect actions} affecting the symbolic monitoring state.% (e.g. by changing some variable local to the DATE). Note that in proper DATEs conditions can depend not just on the monitoring variables' state but also on the program state, however in this paper we limit the analysis to conditions on the monitoring state which is interesting by itself, since unfolding a DATE with such conditions can result in an infinite automaton (e.g. the non-regular language $a^{n}b^{n}$ can be represented with a finite DATE, but not with a finite-state machine).
\footnote{DATEs also include other extensions which we do not deal with in this paper, such as timers and
%, in the same spirit as timed automata \cite{DBLP:journals/tcs/AlurD94}, but extended with stop-watches (e.g. event $c@5$ triggers when timer $c$ reaches $5$ seconds); and (ii) 
 communication channels.
 % for communication between DATEs --- with the actions taken when following a transition possibly involving sending of messages, and starting, stopping, or resetting of a timer. 
 For full semantics of DATEs, refer to \cite{Colombo2009,starvoorsjournal}.} %Even without these extended features, properties in DATEs with conditions and actions are not, in general, amenable to the shadow effect analysis of Clara, since a sequence of Clara-detected NOP shadows that loop in the property can still have an effect on the monitoring state due to transitions' actions.
 
Consider the DATE shown in \ref{f:dateexample}. Transitions are labelled by a triple $e \mid c \mapsto a$ --- when event $e$ occurs and if condition $c$ holds, the transition is taken, executing action $a$\footnote{We leave out the bar and arrow when the condition is true or the action is skip (the identity action).}. For instance, the top transition between states $q_0$ and $q_1$ triggers when a user is whitelisted and the monitoring variable \textit{transferCount} is at least 3, and if taken resets this variable. Applying Clara's first analysis to this property by ignoring the conditions and actions would result in removing the \textit{transfer} transition in state $q_1$ since upon a \textit{transfer} the monitor would never change states. Clearly, taking this transition could have an effect on which future transitions are activated. For similar reasons, Clara's third analysis may disable transitions unsoundly. 

%\gp{We should give an example of a DATE property here (keep it simple --- e.g. 3 bad logins with a counter) for readers to understand. Also use it to explain how it would be be wrongly optimised by Clara. The paragraph below can be then removed.} 

%Clara's analysis are unsound for such DATEs. 
%Consider such an extended transition that is triggered by a certain event, and counts how many times the transition is triggered through a counter, while another transition is conditional upon this count being larger than some number. If the former kind of transition loops at the same state then Clara's analysis would silence any event that may trigger it, since, without conditions and actions, such a shadow would not affect violation, but if it is enabled the events could actually have an effect on violation. This means that Clara's analysis is unsound when lifted naively to DATEs (by simply considering the control-flow of the DATE).

%However, the first two analyses can be easily made sound, as we shall see, while the third analysis is not naively applicable since it characterizes exactly those consecutive shadows that loop between non-violating property states. In this section we shall generalize all of these for DATEs.

%We shall generalize Bodden's results to be applicable to extended automata, in particular to DATEs.

\subsection{Preliminaries}
\begin{figure}[ptb]
\scalebox{0.7}{
\begin{tikzpicture}[every text node part/.style={align=center},node distance=6cm,on grid,auto]
%States
	\node (s) [above = 1 cm of s_0] {};
	\node (ss) [rectangle, draw, fill = gray!30, left = 3 cm of s] {\textit{Initial Variable State:} transferCount = 0;};
	\node (s_0) [circle, draw, initial, initial text={\textbf{For Each:} \textit{User u}}] {$q_0$};
	\node (s_1) [circle, draw] [right = 8cm of s_0]{$q_1$};
	\node (s_2) [circle, draw, accepting, fill=red] [right = 6cm of s_1]{$q_3$};
	\node (s_3) [circle, draw] [below = 2.2cm of s_1]{$q_4$};
	\node (s_4) [circle, draw] [left = 4cm of s_3]{$q_5$};
	\node (s_5) [circle, draw] [below = 2.2cm of s_0]{$q_2$};

%Transitions
	\path[->] [] (s_0)	edge node {$\text{greyList(u)}$} (s_1);
	\path[->] [bend left = 20, below] (s_1)	edge node {$\text{whiteList(u)} \mapsto \text{transferCount} = 0$} (s_0);
	\path[->] [loop above, above] (s_1)	edge node {$\text{transfer(u)}$\\ $\mapsto \text{transferCount}\doubleplus$} (s_1);
	\path[->] [bend right = 20, above] (s_1)	edge node {$\text{whiteList(u)} \mid \text{transferCount} \geq 3$\\ $\mapsto \text{transferCount} = 0;$} (s_0);
	\path[->] [] (s_1)	edge node {\\$\text{whiteList(u)}$\\ $\mid \text{transferCount} < 3;$} (s_2);
	\path[->] [right] (s_1)	edge node {\\$\text{permanentlyDisabled(u)}$} (s_3);
	\path[->] [left] (s_0)	edge node {\\$\text{permanentlyDisabled(u)}$} (s_5);
	\path[->] [above] (s_3)	edge node {$*$} (s_4);
	\path[->] [above] (s_5)	edge node {$*$} (s_4);
\end{tikzpicture}}
	\caption{Example DATE specifying that once a user is greylisted they can only be whitelisted after performing three or more transfers.}
	\label{f:dateexample}
\end{figure}
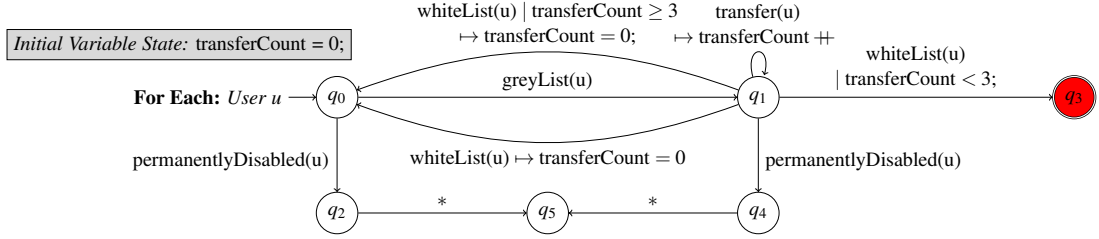
We start by identifying what we mean by a DATE %\footnote{Although DATEs include timer events and are dynamic (multiple DATEs can run in parallel, depending on each other) we do not explore these features in this paper.} 
in this paper, and continue exploring some notions and results we will need to present our residual analysis.

\begin{definition}
A DATE $\D$ is a tuple $\langle \Q, \Sigma, \Theta, \q_{0}, \theta_0, \B, \delta\rangle$, where $Q$ is the set of states, $\Sigma$ is an  alphabet of events, $\Theta$ is the type of monitoring variable states, $q_0\in Q$ is the initial state, $\theta_0 : \Theta$ is the initial monitoring variable state, $\B\subseteq Q$ is a set of bad states, and $\delta \subseteq \Q \times \Sigma \times C \times A \times \Q$ is a transition relation with conditions ($C = \Theta \rightarrow \bool$) and actions ($A = \Theta\rightarrow \Theta$). We write $\q \xrightarrow{\e \mid c \mapsto a} \q'$ for $(\q,\e,c,a,\q') \in \delta$, $\textit{skip}$ for the identity action, and denote the type of DATEs by $\mathbb{D}$.
\end{definition}

Property automata as defined in Section \ref{s:background} can be seen as instances of DATEs with a transition $\q \xrightarrow{\e} \q'$ being translated into $\q \xrightarrow{\e\mid \textit{true}\;\mapsto\; \text{skip}} \q'$.  In our experience with DATEs, the number of states used is typically rather small, and while the symbolic state can be unbounded (e.g. includes lists), in practice such usage is rare and is usually delegated to a database.

%\gp{Nowhere do we give DATE semantics. Perhaps we should. Also, DATEs as we use them here remain in the same state if no transition is triggered, right? This should be said informally (in the example) and also given in the semantics.}\sa{We give it later on when defining satisfaction of ground traces with a DATE no?}

%\cc{can we do without the note about determinism?}
%Determinism of the transition relation of a DATE is typically desirable from a monitoring perspective for efficiency reasons. However, in the presence of actions, determinism is crucial since otherwise it would not be impossible to decide which actions to perform. To ensure determinism, events and conditions on transitions from a state must be mutually exclusive:
%\small\[\q \xrightarrow{\e \mid c \mapsto a} q' \land
%  \q \xrightarrow{\e \mid c' \mapsto a'} q'' \land
%  (\exists \theta \cdot c(\theta) \land c'(\theta)) 
%  \implies
%  c=c' \land a=a' \land q'=q''
%\]\normalsize
%\begin{align*}
%\forall \theta \in \Theta &\cdot \exists (q,e,c,a,q') \in \delta \cdot\\
%&c(\theta) \implies \forall (q,e,c',a',q'') \in \delta \cdot c'(\theta) \implies a = a' \wedge q' = q''%&&(e = e' \wedge c = c' \wedge a = a' \wedge q' = q'')\\
%%&&&\vee \neg c'(\theta)
%\end{align*}
We assume determinism of the transitions: from each state, transitions with the same events have mutually exclusive conditions, with regards to any monitoring variable state. Thanks to this assumption, we can use the transition function in an applicative manner.

\begin{definition}
The concrete transition function of a DATE $\concretedelta \in (Q\times \Theta)\times \Sigma \rightarrow Q\times \Theta$ is defined over the DATE and monitoring variable state:
\small\[\begin{array}{lcl}
\concretedelta((q,\theta), e) &\mydef &
\left\{\begin{array}{ll}
  (q',a(\theta))       & \mbox{ if $q\xrightarrow{e\mid c\mapsto a}q' \land c(\theta)$}         \\
  (q,\theta)  & \mbox{ otherwise}\\
\end{array}\right.
\end{array}\]\normalsize
We will write $\concretedelta^*$
% \in (Q\times \Theta)\times \Sigma^* \rightarrow Q\times \Theta$
 to denote the transitive closure of $\concretedelta$.
%also define the transitive closure of the concrete transition function as follows:
%\[\begin{array}{lcl}
%\concretedelta^*((q,\theta), \langle e\rangle) &\mydef& \concretedelta((q,\theta), e)\\
%\concretedelta^*((q,\theta), \langle e \colon es \rangle) &\mydef& \concretedelta^*(\concretedelta((q,\theta), e), es)
%\end{array}\]
%\begin{align*}
%\delta((q,\theta), e) &\mydef (q',a(\theta)) &&(\exists (q,e,c,a,q') \in \delta \cdot c(\theta))\\
%\delta((q,\theta), e) &\mydef (q,\theta) &&(\nexists (q,e,c,a,q') \in \delta \cdot c(\theta))\\
%\delta^*((q,\theta), \langle e\rangle) &\mydef \delta((q,\theta), e)\\
%\delta^*((q,\theta), \langle e \colon es \rangle) &\mydef \delta^*(\delta((q,\theta), e), es)
%\end{align*}
\end{definition}

This definition makes the DATEs semantics implicitly total with regards to events and conditions, since if there is no transition for a certain event we remain at the same state. %We do not make DATEs explicitly total for conciseness.

Transitioning thus depends on the symbolic state, which we could approximate statically through some kind of code or predicate analysis. However here we shall take the maximal over-approximation, by always considering that a transition's condition may both be true or false. More precise approximations can be relatively more expensive to compute than than this, which is why we made the decision to focus solely on control-flow analysis, leaving the use of predicate analysis for future work. This means that we must consider transitioning into multiple states, which we cater for in the following definition.
%Statically we assume that we cannot know the resolution actions applied to a certain state. Therefore to reason about transitioning statically we define an over-approximation static transition function for DATEs, which takes into account conditions which are syntactically equivalent to true or false, but all others are taken to act completely non-deterministically.

%Since DATEs have conditions and actions statically we need to consider their semantics differently than at runtime, since statically we do not have a program state. Thus statically we must consider all possible program states, and we choose to do this by simply treating a DATE as if it were a non-deterministic automaton, by considering all the possible states immediately reachable from a state given an event, and considering also the possibility that no condition evaluates to true.
% Thus we must consider both the cases that a condition evaluates to true and false. Thus the static transition function of DATEs is defined over sets of states and traces.

%but note that we keep the definition for programs tagged with DATE properties.

\begin{definition}
\label{statictransdef}
Given states $\q,\q'\in Q$ and event $e\in\Sigma$, we say that \emph{$\q$ potentially goes to $\q'$ with event $e$}, written $\q\xrightarrow{e}_{\textit{approx}} \q'$, if a transition with event $e$ and a condition that can be satisfied, or $\q'$ is $q$ and there is no outgoing transition out of it that must be taken on $e$ being triggered:
\small\[\begin{array}{lcl}
  q\xrightarrow{e}_{\textit{approx}} q' & \mydef & 
    (q\xrightarrow{e\mid c\mapsto a} q' \land c\neq \textit{false}) \lor    (q=q' \land \not\exists q'' \cdot q''\neq q \land q\xrightarrow{e\mid \textit{true} \mapsto a} q'')
\end{array}\]\normalsize

Given a DATE, the \emph{static transition function} $\approximatedelta_\D \in 2^Q\times \Sigma \rightarrow 2^Q$ is defined to be the function which, given a set of states and an event, returns the set of states potentially reachable from any of the input states:  $\approximatedelta_\D(S,\e)\mydef \{ q' \mid \exists q\in S \cdot q \xrightarrow{e}_{\textit{approx}} q'\}$.

We use $\approximatedelta^*_\D \in 2^Q\times \Sigma^* \rightarrow 2^Q$ to denote its transitive closure
%, also writing $\Q \xRightarrow{t}_{\D} \Q'$ to denote that $\approximatedelta^*_\D(Q,t)=Q'$
. Finally, we use $q \hookrightarrow_{\D} q'$ to denote that $q'$ is reachable from $q$ with the static transition relation:
$q \hookrightarrow_{\D} q' \mydef \exists t \cdot q' \in \approximatedelta^*_\D(\{q\}, t)$.
%
%\begin{align*}
%q \hookrightarrow_{\D} q' \mydef \exists t : \Sigma^*_\D \cdot q' \in \approximatedelta^*_\D(\{q\}, t)
%\end{align*}
\end{definition}

\noindent We can prove that given a state and a ground trace, then the state reached by $\concretedelta^*$ is also possibly reached (with $\approximatedelta^*$) from that state, with that trace, i.e. $\approximatedelta^*$ over-approximates $\concretedelta^*$.

%\begin{theorem}
%\label{statictransapproximatesstep}
%The approximate transition function is an over-approximation of the concrete transition function:
%
%$\forall \theta : \Theta, t : \Sigma^*, q \in \Q \cdot \exists \theta' : \Theta \cdot \concretedelta((q,\theta'),t) = (q',\theta) \implies q' \in \approximatedelta(\{q\},t)$.
%\end{theorem}

\begin{theorem}
\label{statictransapproximates}
%The transitive closure of the approximate transition function is an over-approximation of the transitive closure of the concrete transition function:
%
\small$\forall \theta : \Theta, t : \Sigma^*, q \in \Q \cdot \exists \theta' : \Theta \cdot \concretedelta^*((q,\theta'),t) = (q',\theta) \implies q' \in \approximatedelta^*(\{q\},t)$\normalsize.
\end{theorem}

As we did for property automata, we define what it means for a DATE to satisfy the different kinds of traces using this static approximation of transitioning at runtime, and relate it to the static transition function.%\gp{There should be exact semantics of DATEs taking $\theta$ into account. The previous theorem can then be used to show that exact satisfaction implies approximate satisfaction. Also, keep in mind that the system will change $\theta$ between events.}

\begin{definition}
%A parametrized static trace $\st$ is said to satisfy a DATE $\D$ if for each object $\ob$ the corresponding ground traces never contain a trace that when applied to the property results in a bad state.
%The satisfaction operator for the different kinds of traces is defined similarly for DATEs, using the transitive closure of a DATE.
A ground trace $t \in \Sigma^*$ is said to satisfy a DATE $\D$ (with $\Sigma \subseteq \Sigma_\D$), if none of its prefixes applied to the transitive closure of $\D$, starting from the initial state of $\D$ and the initial monitoring variable state, lead to a bad state of $\D$: $t \vdash \D \mydef \forall t' \in \textit{prefixes}(t) \cdot \concretedelta^*((q_0,\theta_0),t') = (q, \theta) \wedge q \not\in \B_\D$.

%\begin{align*}
%t \vdash_\theta \D &\mydef \forall t' \in \textit{prefixes}(t) \cdot \exists \theta' : \Theta \cdot \concretedelta^*((q_0,\theta),t') = (q, \theta') \wedge q \not\in \B_\D
%\end{align*}
%
%To be able to take about all possible runs of a program, we extend this to range over all possible start states of a program, $\Theta_{\textit{start}} \subseteq \Theta$:
%\begin{align*}
%t \vdash \D &\mydef \forall \theta \in \Theta_{\textit{start}} \cdot t \vdash_\theta \D
%\end{align*}
%\begin{align*}
%t \vdash \D &\mydef \forall t' \in \textit{prefixes}(t) \cdot \concretedelta^*((q_0,\theta_0),t') = (q, \theta) \wedge q \not\in \B_\D
%\end{align*}
We define the satisfaction of a parametrised trace and a statically parametrised trace as before with respect to a DATE and over this ground trace satisfaction operator, and similarly for the equivalence relations between DATEs.% Similarly, we define the equivalence relations between DATEs and programs using this satisfaction with respect to a DATE.%\gp{Which other satisfaction operators? I do not see what you mean by this sentence.}
\end{definition}

%We can also relate satisfaction of a ground trace with a DATE with respect to the static transition function.

Base on the previous theorem, we can show satisfaction of a trace by considering whether any of its prefixes possibly lead to a bad state.

\begin{theorem}
\label{satisfactionwithapproximatefunction}
If the static transition function applied to any prefix of $t$ from the start state of $\D$ does not contain a bad state, then $t$ satisfies $\D$: (0$\forall t \in \Sigma^*\cdot \forall t' \in \textit{prefixes}(t) \cdot \approximatedelta(\{q_0\},t) \cap \B_\D \neq \emptyset) \implies t \vdash \D$.
\end{theorem}
Moving on to creating residuals, from \ref{f:dateexample} we can see that not all states are useful: consider how being at $q_2$, $q_4$ and $q_5$ implies that one can no longer violate the property. $q_5$ can however only be reached through $q_2$ or $q_4$ where the monitor verdict is clear, and thus $q_5$ is useless. Thus we should keep states $q_2$ and $q_4$ (since before reaching them we do not know the verdict of the trace), but we do not need $q_5$ (since satisfaction is clear). We then classify a state as useful if: (1) it is reachable from the initial state, and it can reach a bad state ($q_1$), or (2) it is reachable by one step from a state of the first kind ($q_2$ and $q_4$).

\begin{definition}
A state $q$ in DATE $\D$ is said to \emph{possibly lead to a violation in $\D$}, written $\textit{badAfter}(q)$, if it is reachable from the initial state and a bad state is reachable from it:
\small$\textit{badAfter}(q) \mydef 
    q_0 \hookrightarrow q \land 
    \exists q' \in B \cdot q \hookrightarrow q'
$\normalsize.

A state $q$ in DATE $\D$ is said to be \emph{an entry-point to a satisfied region in $\D$}, written $\textit{goodEntryPoint}(q)$, if it cannot possibly lead to a violation and is one transition away from such a state that can possibly lead to a violation:
\small$
\textit{goodEntryPoint}(q) \mydef 
    \neg\textit{badAfter}(q) 
    \land \exists q' \cdot q' \xrightarrow{e}_{\textit{approx}} q \land \textit{badAfter}(q')
    % \xrightarrow{e\mid c \mapsto a} q \land \textit{badAfter}(q')
$\normalsize.

A state $q$ in DATE $\D$ is said to be \emph{useful in $\D$}, written $\textit{useful}(q)$, if it can possibly lead to a violation or is an entry-point to a satisfied region in $\D$: 
\small$
\textit{useful}(q) \mydef 
    \textit{badAfter}(q) \lor \textit{goodEntryPoint}(q)
$\normalsize.
Given a DATE $\D$, it can be reduced to the reachable useful states to obtain $\reachable{\D}$ which contains only states in $\D$ which are $\textit{useful}$, and the transitions between them: \small$\delta_{\reachable{\D}} \mydef \{(\q,\e,c,a,\q') \in \delta \mid \textit{useful}(\q) \wedge \textit{useful}(q')\}$\normalsize.%\gp{Adapt proofs accordingly}
%\begin{align*}
%\reachable{\delta} &\mydef \{d = (\q,\e,c,a,\q') \in \delta \mid \q_0 \hookrightarrow \q \\
%&\qquad\qquad\qquad\qquad\qquad\qquad \wedge (\textit{sideEffect}(d) \\
%&\qquad\qquad\qquad\qquad\qquad\qquad\qquad \vee \textit{effectAfter}(\q')\\
%&\qquad\qquad\qquad\qquad\qquad\qquad\qquad \vee \exists \q_b \in \mathbb{\B} \cdot \q \hookrightarrow \q_b\}% \vee \q' \hookrightarrow \q_b)\}%\\
%%\delta_{\reachable{\D}} &\mydef \reachable{\delta} \cup \{(\q,\e,\textit{true},\textit{id}_A,\q) \mid \nexists \q' \cdot (\q,\e,c,a,\q') \in \reachable{\delta}\}
%\end{align*}

%$\reachable{\D}$ is then $\D$ with $\reachable{\delta}$ as its transition function, without the states not in $\reachable{\delta}$.
\end{definition}

%Note that we extend the reachability-reduced transition function s.t. it is total with respect to events, i.e. there is always a transition from each state with each event in the alphabet. However it is not total with respect to conditions explicitly, but the static transition function interprets the DATE as if, for each event $\e$, there was a $\e$-transition into the same state with the negated disjunction of every other condition corresponding to an $\e$-transition out of the same state.

%We can also extend the notion of a side-effect between transitions.
%
%\begin{definition}
%A transition $t$ in a DATE $\D$ is said to have an effect on a transition $t'$ ($t \triangleright t'$) if its action has a side-effect, or if some other transition reachable from it has a global side-effect. 
%\begin{align*}
%\textit{globalEffect}((\q,\e,c,a,\q')) &\mydef \textit{sideEffect}(a) \vee \exists (\q',\e',c',a',\q'') \in \delta_{\D} \cdot \textit{sideEffect}_{\D}(t')
%\end{align*}
%\end{definition} 

%Note that the semantics of accepting states in DATEs are such that when at runtime an accepting state is reached then the monitor is garbage collected. 

We can then show that a DATE reduced for reachability is equivalent to the original DATE with respect to an approximation of a program, by using the theorems and propositions presented in the previous sections, which we claim also apply to DATEs by simply using the DATE satisfaction operators.%\gp{Why is this using the approximate semantics? Shouldn't we be proving that under the exact semantics of the DATE reachability reduction retains the same meaning of a DATE?}\sa{Because before I defined the semantics of DATEs w.r.t. to the approximate transition function :P anyway now I defined it with the concrete function; but with Theorem 2. we use the approximate function too. So it's fine to leave it this way.}

\begin{theorem}
\label{reachabilitytheorem}
A DATE $\D$ is equivalent to its reachability-reduced counterpart $\reachable{\D}$ (with alphabet $\Sigma$), with respect to any set of traces:
$
\forall T \subseteq \Sigma^* \cdot \D \cong_T\reachable{\D}
$.
\end{theorem}

%From this it follows easily that the two DATEs are equivalent with respect to a static approximation of a program.
%\begin{corollary}
%A reachability-reduction is equivalent to the original DATE.
%\label{reachabilitycor}
%\begin{align*}
%\D \cong_{P^S_{\Sigma}} \reachable{\D}
%\end{align*} 
%\end{corollary}
%Note that if formally we were considering parallelly executing DATEs we could also take into account equivalence between sets of parallelly executing DATEs.\\

%\gp{This should be in a separate subsection, with better explanations. As it is, it is almost just defns, theorems, with minor text between. The section should start with the overall picture that is achieved in the section, then give the results after that.}

\subsection{Residual Analysis}

We are concerned with producing \textit{residuals} of a DATE \cite{isolares} with respect to some known information about the program --- the part of the property which cannot be proved from what we know about the program. Recall that previously we informally characterized a residual $D'$ of a DATE $D$, given a static program $P^S_\Sigma$ over that same alphabet, by one property: the intersection of the program intersected with that of property and separately with the residual are equal: where $L(D)$ is the set of bad traces accepted by $D$, $L(D') \cap (P^S_\Sigma \Downarrow \OId) = L(D) \cap (P^S_\Sigma \Downarrow \OId)$. This is equivalent to our notion of equivalence: $D' \cong_{P^S_\Sigma} D$. Note that the reachability-reduction just defined is a residual for all programs.

%is a subset of the possible violating behaviour of a property with respect to a program. Residuals can thus be reductions of a property that remain equivalent to the original property with respect to some behaviour.
In this section, we start by presenting a number of definitions and results which we shall use to extend the analysis used in Clara to create DATE residuals.
%, results which allow us to reason about when the reduction of DATEs with respect to some alphabet is equivalent to the original DATE, in the context of some program's behaviour. We also introduce the notion of union between two reductions of a DATE, to allow the behaviour of multiple objects to be reasoned about individually but unified for the monitor at runtime.

One of the ways we shall be creating residuals is by restricting the alphabet of a DATE, the result of which is equivalent to the original DATE, with respect to some kind of traces.

\begin{theorem}
\label{subdatethm}
A DATE $\D$ with alphabet $\Sigma$ and its alphabet-restriction by some alphabet $\Sigma' \subseteq \Sigma$ are equivalent with respect to a set of ground traces $T \subseteq \Sigma^*$:
$
\D \upharpoonright \Sigma(T) \cong_T \D
$
\end{theorem}

Since we are looking at parametrised traces, different objects in a program activate different instances of the monitor. These objects may have different behaviour and thus they may use different subsets of the DATE alphabet. We will thus, in the next section, consider the residuals of a DATE with respect to different objects, however each of these residuals individually is not enough to monitor the whole program soundly with. We define a union operator over DATEs to allow this. 

To ensure that the combination of DATEs ($\D$ and $\D'$) remains a DATE we require that they start from the same initial state and monitoring state, and that a DATE exists that captures both of their behaviour: $\exists D'' : \mathbb{D} \cdot \Q_{{\D}} \cup \Q_{{\D}'} \subseteq \Q_{{\D}''} \wedge  \Sigma_{\D} \cup \Sigma_{{\D}'} \subseteq \Sigma_{{\D}''} \wedge {\B}_{{\D}} \cup {\B}_{{\D}'} \subseteq {\B}_{{\D}''} \wedge \delta_{{\D}} \cup \delta_{{\D}'} \subseteq \delta_{{\D}''}$, i.e. $\D$ and $\D'$ are said to be \emph{component-wise subsets} of $\D''$.% that are component-wise subsets of another DATE,\cc{not sure i understood what are component-wise subsets... why not provide the formal def?} otherwise it cannot be assured that the union produces a DATE (determinism may be lost).

\begin{definition}
\label{uniondef}
Given two DATES that are component-wise subsets of another DATE\footnote{This ensures that the union, as defined here, is in turn merely the bigger DATE without some states and/or transitions, which is still a DATE, since determinism is preserved.}, their component-wise union is defined as the property with both of their transitions, states and bad states, and starting from the same initial state:
\small${\D} \sqcup {\D}' \mydef \{\Q_{{\D}} \cup \Q_{{\D}'}, \Sigma_{\D} \cup \Sigma_{{\D}'}, \q_{0}, \theta_0, {\B}_{{\D}} \cup {\B}_{{\D}'}, \delta_{{\D}} \cup \delta_{{\D}'} \}$\normalsize
\end{definition}

%\begin{proposition}
%Given a DATE $\D$, the component-wise union of two sub-DATES $\D'$ and $\D''$, of $\D$, is itself a sub-DATE of $\D$.
%\begin{align*}
%\D' \structsubseteq \D \wedge \D'' \structsubseteq \D \implies \D' \sqcup \D'' \structsubseteq \D
%\end{align*}
%\end{proposition}
%\begin{proof}
%This follows immediately from Defn. \ref{subDatedef} and Defn. \ref{uniondef}.
%\end{proof}

%%does not work
%\begin{proposition}
%Given a DATE $D$, and a sub-DATE of it $D'$, then if a trace $T$ satisfies the alphabet-reduction of $D$ by the alphabet of $T$, then  it also satisfies the union of the reachable-reductions of the sub-DATEs.
%\begin{align*}
%\forall t : T \cdot t \vdash D \upharpoonright \Sigma(T) \implies t \vdash D \upharpoonright \Sigma(T) \cup \D'
%\end{align*}
%\end{proposition}
%\begin{proof}
%The reachable-reduction of a DATE maintains only the states reachable from the initial state, making sure that the union does not recognize any language not recognized by both automata.
%\end{proof}

We shall be producing residuals of a DATE by using the alphabet-restriction, then the reachability-reduction, and then performing the union on all these residuals. The next theorem shows that this union preserves the behaviour of the single reduced DATEs.

\begin{theorem}
\label{2alphredequivimpliesunionequiv}
Given sets of traces $T_0, T_1 \subseteq \Sigma^*$, and DATE $\D$, the union of $D$'s alphabet-restriction with respect to each of the sets of traces is equivalent to $\D$ with respect to the union of the sets of traces:
\small$
\reachable{\D \upharpoonright \Sigma(T_0)} \sqcup  \reachable{\D \upharpoonright \Sigma(T_1)} \cong_{T_0 \cup T_1} \D$\normalsize.
\end{theorem}

We can now move on to presenting our constructions of residuals.

\subsection{Residual Constructions}
%\gp{Weird how there's such a big chunk before section 3.1. Reorganise.}

%As opposed to Clara, where the analyses are plugged in after the instrumentation process, to turn off certain instrumentation points, we want our analysis to be more general. 
We start by formally describing analyses which both prune the property by removing transitions and states that are irrelevant for the program's violation, and silence statements that the analysis concludes will not affect violation. 
We present three residuals, one without the events used by the program, another taking into account whether events can occur on the same object, and the last one removing from the DATE transitions that can never be used by a trace in the program. Theoretically, using the last analysis is enough, since each analysis is finer than the other, however in practice one may want to first use the other analyses since they are cheaper to compute. We prove that each of these is equivalent to the original DATE with respect to the given program. The last two analyses also provide the opportunity of identifying statements in the program that can be silenced safely.

%\subsubsection{Pre-Analysis} Before residual analysis of a property with a program we analyse the property by itself to remove any useless states and transitions, thus for the analyses we present we assume that: (1) we reduce the property with the reachability analysis; and (2) there are no looping transitions at the same state that do not have a side-effect.  
%%and (3) there is no state from which a bad state (accepting or bad) is not reachable. 
%%Currently we define a transition to not have a side-effect if it does not have an action, however we will make this less strict later.
%%\begin{definition}
%%An action is said not to have any side effect if it is the empty action.
%%\begin{align*}
%%\textit{sideEffect}(a) &\mydef a \neq \emptyset%\\
%%%\textit{sideEffect}((\q,a,\textit{cond},\textit{act},\q')) &\mydef \textit{sideEffect}(a)
%%\end{align*}
%%\end{definition} 
%
%%Note that DATEs have a semantics such that looping transitions not need to be explicit. 
%
%
%From now on we assume these conditions hold for each \textit{DATE} we consider.
 
\subsubsection{Absent event pruning} Recall that Clara's first analysis computes the symbols that should be monitored, thus excluding: (i) events that appeared only on transitions looping in the same state, (ii) symbols that do not appear in the program, and (iii) symbols only outgoing from states from which a bad state is not reachable in the property reduced by the previous types of symbols. We now consider these in the case of DATEs. 

We cannot remove transitions such as in (i) since such idempotent transitions may perform actions which affect the triggering of other transitions (consider the looping transition on state $q_3$ in \ref{f:dateexample}). Those of type (ii) can be removed safely, since if a certain symbol does not appear in the program, then clearly transitions tagged by such symbols in the property will never be taken and can be removed. While our reachability-reduction already takes care of events of type (iii)\footnote{Note that this would not be always safe if we were considering multiple DATEs executing at the same time.}. %Thus, as with (i) we can perform a reachability analysis, but only remove transitions without side effects. So, if a state cannot reach a final state, consider each of its outgoing transitions, if such a transition does not have a side effect and transitions to another state then we remove it if the other state does not have side effects after the same type of analysis.% (we call the transition function reduced this way $\reachable{\reachable{\delta}}$). 
%also applies, while also extending it to apply for accepting states, since in Clara we had an accepting semantics while in \textit{DATE}s accepting states are explicit. We can use the reduction operator to remove the applicable transitions of type (ii), similarly the transitions of type (iii) can be removed by simply performing a reachability analysis on the property states.
We thus define $\textit{residual}_0$ over a property ${\D}$, with respect to a program $P^S_\Sigma$:
\small$
\textit{residual}_{0}({\D}) \mydef \reachable{{\D}\upharpoonright {\Sigma(P^S_\Sigma)}}
$\normalsize.

Based on Thm. \ref{reachabilitytheorem}, Thm. \ref{subdatethm}, and Prop. \ref{staticprogesatequivtoprojectionsat2}, we can show that this is equivalent to the original DATE with respect to the used program.
%\begin{proposition}
%\begin{enumerate}
%\item The reachable reduction of a property is a sub-property of the property.
%\item The alphabet-reduction of a property is a sub-property of the property.
%\item The reachable reduction of the alphabet-reduction of a property is a sub-property of the property.
%\end{enumerate}
%\end{proposition}

\begin{theorem}
The $\textit{residual}_0$ of a DATE $\D$ is equivalent to $\D$, with respect to program approximation $P^S_\Sigma$ (with $\Sigma \subseteq \Sigma_{\textit{residual}_0(\D)}$): \small $\textit{residual}_{0}({\D}) \cong_{P^S_\Sigma} {\D}$\normalsize
\end{theorem}
%\begin{proof}
%$\textit{residual}^P_{0}({\D})$ is clearly a sub-DATE of $\D$, so by using Prop. \ref{subdateprop} we are left to prove that $P^S_\Sigma \Vvdash \textit{residual}^P_{0}({\D}) \implies P^S_\Sigma \Vvdash \D$:\\
%\begin{align*}
%&P^S_\Sigma \Vvdash \textit{residual}^P_{0}({\D})\\
%&\implies (\textit{by definition of } \Vvdash)\\
%&\Sigma \subseteq \Sigma_{\textit{residual}^P_{0}({\D})}\\
%&\implies (\textit{since } \textit{residual}^P_{0}({\D}) \structsubseteq \D)
%&\Sigma
%&\forall \st \in P^S_\Sigma \cdot \forall \oid \in OId \cdot \forall t \in \st \Downarrow_{\may} \oid \cdot t \vdash \textit{residual}^P_{0}({\D})\\
%&\implies (t \vdash \textit{residual}^P_{0}({\D}) \implies \Sigma(t) \subseteq \Sigma_{\textit{residual}^P_{0}({\D})}\\
%&\forall \st \in P^S_\Sigma \cdot \forall \oid \in OId \cdot \forall t \in \st \Downarrow_{\may} \oid \cdot t \vdash \D\\
%&\implies \\
%&P^S_\Sigma \Vvdash \D
%\end{align*}
%\end{proof}

\subsubsection{Object-specific absent event pruning} 
%\cc{i don't really like this title but not sure how to improve hehe}

%\sa{mention more about shadows and turning them off in background}
Like Clara, we can generalise the first analysis to consider events occurring on objects. Given an object, if the traces corresponding to it do not use the full alphabet of the DATE, then we can create a residual that is enough to monitor just that object soundly. Consider \ref{f:dateexample}, if we have an object that we know is never greylisted but performs transfers then we can keep only the transitions triggered upon a transfer. Based on the results given in Thm. \ref{reachabilitytheorem} and Thm. \ref{subdatethm}, we define this as:
\small$
\textit{residual}_{1}({\D},\oid) \mydef \reachable{{\D}\upharpoonright {\Sigma(P^{S}_\Sigma \Downarrow \oid)}}$\normalsize.

%\cc{earlier we were using ObjId right? Why did we switch to oid here?}
\begin{proposition}
\label{equivtoobjprojection}
The $\textit{residual}_1$, for an identifier $\oid$, is equivalent to the original DATE $\D$ with respect to the traces of $\oid$ in the program $P^S_\Sigma$:
\small$
\textit{residual}_{1}({\D},\oid) \cong_{P^S_\Sigma \Downarrow \oid} \D$\normalsize.
\end{proposition}

For runtime monitoring we now have two choices: (1) create a different monitor for each object identifier, corresponding to the associated residual, such that the monitor only instruments the statements associated with the identifier; (2) perform the union on all the residual DATEs and instrument the program as usual (i.e. by simply matching the DATE events with statements). The former would require new instrumentation techniques that employ static aliasing knowledge, while with the latter one could use existing techniques. 
%\[\textit{residual}^P_1({\D}) \mydef \bigsqcup_{\oid \in \OId} \textit{residual}^P_1({\D}, \oid)\]
Using Prop. \ref{equivtoobjprojection}, Prop. \ref{2alphredequivimpliesunionequiv} and  Prop.  \ref{staticprogesatequivtoprojectionsat2}, we can show that the resulting DATE in the second choice is still equivalent to the original one, with respect to the program.

\begin{theorem}
Given the residual for each identifier, then their union is equivalent to the original DATE, with the program approximation:
%The union of each residual of a DATE $\D$, associated with each object identifier, is equivalent to $\D$ with respect to the static approximation $P^S_\Sigma$.
\small$\bigsqcup_{\oid \in \OId} \textit{residual}_1({\D}, \oid) \cong_{P^S_\Sigma} {\D}$\normalsize.
\end{theorem}
%\begin{proof}
%By Prop. \ref{equivtoobjprojection} and Prop. \ref{2alphredequivimpliesunionequiv} clearly the DATEs are equivalent with respect to the ground traces generated by the static approximation of the program, and using Prop.  \ref{staticprogesatequivtoprojectionsat2} the theorem follows.
%\end{proof}

Consider again an object that only performs transfers. With respect to this, \ref{f:dateexample} would be reduced to just the initial state using the second residual, and thus we can statically conclude that the object satisfies the property. In this manner, we no longer need to monitor this particular object when it performs a transfer, and thus any transfers associated solely with it can be silenced. Using an object-specific residual, we can then reduce the traces in the static program by removing identifier-events pairs, where the event does not appear in the identifier's residual: \small$\textit{noEffect}((\oid, e), \D) \mydef \e \not\in \Sigma_{\textit{residual}_i(\D,\oid)}$\normalsize. 

A static program without such pairs will remain equivalent to the original static program, with respect to that residual (i.e. corresponding traces in the respective programs will maintain the same verdict with respect to the residual)\footnote{See \cite{technicalreport} for a full formalisation of what this means, and a proof for that statement, which we have omitted from here, given lack of space.}.

\subsubsection{Unusable transition pruning} In the previous two analyses we have ignored the flow of events, in fact we consider only the alphabet of the program and DATE. %, which makes these analyses relatively cheap to compute. 
Our novel third analysis, however, makes use of the control-flow, by considering the possible traces of a program. 

Consider \ref{f:dateexample}, and the trace $\textit{whitelist} ; \textit{greylist} ; \textit{transfer}^n$ as the program (over a single object). Given the previous two analyses, only transitions between states $q_0$, $q_1$ and $q_3$ would remain. However, the \textit{whitelist} transitions from $q_1$ can never be activated since the trace only performs \textit{whitelist} before the user is greylisted. Thus we can remove these transitions. We define what it means for a trace to use a transition.

\begin{definition}
A trace is said to use a transition from $q$ to $q'$ with an event $e$, in a DATE $\D$, if the transitions condition is not false, and a strict prefix of the trace can potentially go to $q$ and the head of the remaining suffix is $e$:
\small$\textit{uses}(t, q \xrightarrow{e \mid c \mapsto a} q') \mydef c \neq \textit{false} \wedge \exists t' \doubleplus \langle e \rangle \in \textit{prefixes}(t)\cdot q \in \approximatedelta^*_\D(\{\q_0\}, t')$\normalsize
\end{definition}

In general, given a trace, this does not always trigger all the transitions in a DATE, in fact we can prune away the unused transitions and the residual DATE will remain equivalent to the original one with respect to that trace. We can generalise this notion by removing transitions that cannot be used by any of the traces: \small$\textit{residual}^P_2(\D, P^S_\Sigma) \mydef \reachable{\D \upharpoonright \{d \in \delta \mid \exists t : P^S_\Sigma \Downarrow \OId \cdot \textit{uses}(t, d)\}}$\normalsize.\footnote{Note that we can also define this for singular objects, $\textit{residual}^P_2(\D, P^S_\Sigma, \oid) \mydef \reachable{\D \upharpoonright \{d \in \delta \mid \exists t : P^S_\Sigma \Downarrow \oid \cdot \textit{uses}(t, d)\}}$ and perform the same $\textit{noEffect}$ analysis as with the previous residual.}

%\small\begin{align*}
%%\textit{residual}^P_2(\D, P^S_\Sigma) \mydef \D \upharpoonright \{\delta_i : \delta_\D \mid &% \neg \textit{sideEffect}(\delta_i)\\&   \wedge 
%%\forall t : P^S_\Sigma \Downarrow \oid, pt \in \textit{prefixes}(t) \cdot\\
%%& \qquad\qquad  \approximatedelta^*_\D(\{\q_0\}, pt) \neq \approximatedelta^*_{\D'}(\{\q_0\}, pt)\}
%%\textit{residual}^P_2(\D, P^S_\Sigma) \mydef \D \upharpoonright \{q \xrightarrow{e \mid c \mapsto a} q' \mid &\forall t : P^S_\Sigma \Downarrow \oid, t' \doubleplus \langle e' \rangle \in \textit{prefixes}(t) \cdot\\
%%& \qquad\qquad q \in \approximatedelta^*_\D(\{\q_0\}, t') \implies e' \neq e
%%\textit{residual}^P_2(\D, P^S_\Sigma) \mydef \reachable{\D \upharpoonright \{q \xrightarrow{e \mid c \mapsto a} q' \mid &\exists t : P^S_\Sigma \Downarrow \OId, t' \doubleplus \langle e' \rangle \in \textit{prefixes}(t)\cdot q \in \approximatedelta^*_\D(\{\q_0\}, t') \land e' = e\}}
%%\textit{residual}^P_2(\D, P^S_\Sigma) \mydef \reachable{\D \upharpoonright \{q \xrightarrow{e \mid c \mapsto a} q' \mid &\exists t : P^S_\Sigma \Downarrow \OId, t' \doubleplus \langle e \rangle \in \textit{prefixes}(t)\cdot q \in \approximatedelta^*_\D(\{\q_0\}, t')\}}
%\textit{residual}_2(\D, P^S_\Sigma) \mydef \reachable{\D \upharpoonright \{d \in \delta \mid &\exists t : P^S_\Sigma \Downarrow \OId \cdot \textit{uses}(t, d)\}}
%\end{align*}\normalsize
%%\end{definition}

\begin{theorem}
\label{controlflowresthm} The $\textit{residual}_2$ is equivalent to the original DATE: $\textit{residual}_2(\D, P^S_\Sigma) \cong_{P^S_\Sigma} \D$
\end{theorem}

Previously we discussed informally that multiple CFGs that approximate the program's behaviour can be created, meaning we may have multiple over-approximations of a program, we can thus apply the $\textit{residual}_2$ on each of these successively, creating a possibly finer residual than we can with one over-approximation.
\small\[\begin{array}{lll}
\textit{residual}_2(\D, \langle \rangle) &\mydef \D\\
\textit{residual}_2(\D, P^S_\Sigma : \textit{Ps}) &\mydef \textit{residual}_2(\textit{residual}_2(\D,P^S_\Sigma), \textit{Ps})
\end{array}\]\normalsize

%\begin{theorem}
%\begin{align*}
%D \equiv_{\bigcap_{P^S \in \textit{Ps}} P^S \Downarrow \textit{Oid}} \D
%\end{align*}
%\end{theorem}

\begin{theorem}
\label{controlflowres2thm}
Given a set of static program over-approximations, then applying the third residual construction, consecutively, returns one which is equivalent to the original with respect to the intersection of projection into ground traces of each approximation: \small$\textit{residual}_2(\D, \textit{Ps}) \cong_{\bigcap_{0 < i < \textit{length}(\textit{Ps})} \textit{Ps}(i) \Downarrow \OId_{i}} \D$\normalsize \footnote{This is significant because the ground traces generated by static programs are always a subset of that generated by the runtime program, ensuring that the intersection of the statically generated traces here is a subset of those at runtime.}
\end{theorem}
%\begin{proof}
%This follows from Thm. \ref{controlflowresthm}, Prop. \ref{staticprogesatequivtoprojectionsat2}, and Prop. \ref{equivintersection}.
%\end{proof}

%Note how Thm. \ref{controlflowres2thm} implies that the residual and DATE are equivalent to the program's behaviour at runtime, since by Assump. \ref{staticassump} each over-approximation's projection contains the runtime program..
%
%\begin{align*}
%\textit{ips} = \{\textit{ip} ; \IP \mid \forall t : P^S_\Sigma \cdot &\nexists t_1 \in \textit{prefixes}(t) \cdot \\
%&\exists t_2 \in \textit{prefixes}(t_1) \cdot \\
%&\quad \approximatedelta^*(\D, t_1) = \approximatedelta^*(\D,t_2)\\
%&\quad \wedge \textit{sideEffectAfter}(\approximatedelta^*(\D, t_1), t_1 - t_2)\}\\
%&\quad \wedge 
%\end{align*}

Clara's third analysis silenced statements in the program that together do not have any effect on the flow with respect to the property. However, given conditions on transitions in DATEs, we do not necessarily know if a transition will be activated or not. Hence, the static transition of a DATE ranges over a set of states, considering both possibilities of a transition being taken or not. Using the same principle, we can apply Clara's third analysis to method invocations that only trigger DATE transitions without conditions (or rather with the \textit{true} condition), and that do not have actions with side-effects. However, we do not detail this here given space constraints.

\section{Case Study}
\label{s:case-study}

\begin{figure}
\scalebox{0.7}{
  \begin{tikzpicture}[every text node part/.style={align=center},node distance=6cm,on grid,auto]
	%States
	%start
	\node (startbronze) [initial, initial text = {\textit{Bronze}\\\textit{UserInfo}}, accepting, circle, draw] [] {$\q_0$};
	\node (createdbronze) [circle, accepting, draw] [right = 3cm of startbronze]{$\q_1$};
	\node (activatedbronze) [circle, draw] [right = 3cm of createdbronze]{$\q_2$};
%
%	Transitions
%	
	\path[->] [] (startbronze)	edge node {\textit{createdUser}} (createdbronze); 
	\path[->] [] (createdbronze)	edge node {\textit{activate}} (activatedbronze); 
	\path[->] [loop above] (createdbronze)	edge node {\textit{activate}} (createdbronze); 
	\path[->] [bend left = 20] (activatedbronze)	edge node {\textit{blacklist,greylist,whitelist}} (createdbronze); 
	\path[->] [loop above] (activatedbronze)	edge node {$\tau$} (activatedbronze); 
%
	%States
	%start
	\node (startsilver) [initial, initial text = {\textit{\{Gold/Silver\}}\\\textit{UserInfo}}, right = 4cm of activatedbronze, accepting, circle, draw] [] {$\q_0$};
	\node (createdsilver) [circle, accepting, draw] [right = 3cm of startsilver]{$\q_1$};
	\node (activatedsilver) [circle, draw] [right = 3cm of createdsilver]{$\q_3$};
	%blacklistedlowrisk
	\node (whitelistedsilver) [circle, draw] [below = 1.1 of activatedsilver]{$\q_4$};
	\node (blacklistedsilver) [circle, draw] [right = 3cm of activatedsilver]{$\q_5$};
%
%	Transitions
%	
	\path[->] [] (startsilver)	edge node {\textit{createdUser}} (createdsilver); 
	\path[->] [] (createdsilver)	edge node {\textit{activate}} (activatedsilver); 
	\path[->] [loop below] (createdsilver)	edge node {\textit{activate}} (createdsilver); 
	\path[->] [] (activatedsilver)	edge node {\textit{blacklist}} (blacklistedsilver); 
	\path[->] [] (activatedsilver)	edge node {\textit{whitelist}} (whitelistedsilver); 
	\path[->] [] (whitelistedsilver)	edge node {\textit{pay}} (createdsilver); 
	\path[->] [bend right = 30] (blacklistedsilver)	edge node {$\tau$} (createdsilver); 
	\end{tikzpicture} 	
}
	\caption{CFG lifted with respect to aliasing of each sub-type of UserInfo, with respect to the property in \ref{fig:fitsproperty}.}
  \label{fig:cfgs}
\end{figure}
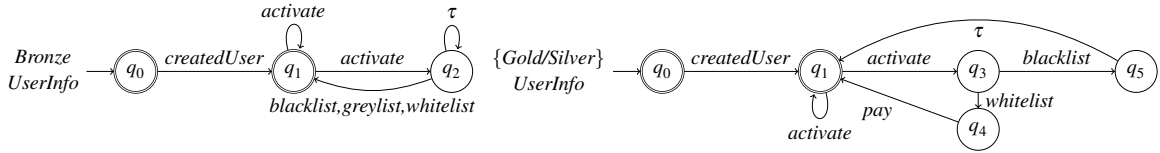

These analysis techniques presented in this paper have been implemented in a tool\footnote{The tool can be downloaded from https://github.com/shaunazzopardi/clarva.}. It is worth noting that due to the differences in the analysis techniques, our tool does not build directly on Clara, although it uses the Soot \cite{soot} tool for Java program analysis. The tool was evaluated them on a simple financial transaction system with users connecting to a proxy in order to enable communication with a transaction server inspired by the industrial systems we have previously used runtime verification on \cite{opesefm}. The proxy and the transaction server can be two different services (the client and provider), possibly provided by different providers, with the property specifying the behaviour that the transaction server expects out of the proxy. 
 %In this way the case study emulates two services with one imposing on the other a certain contract to be allowed access to it. More abstractly, this financial transaction system simulates users effecting payment transactions with some companies, through payment cards. 

In order to see how our approach scales with increased system load and monitoring overhead, the system was evaluated with different numbers of users behaving in a controlled random manner, thus allowing for repeatability of the experiment. The static analysis was performed three times: (1) with the original property to produce the program monitored by the first residual; (2) with the first residual to produce the program monitored by the second residual property, with some method calls silenced according to the instrumentation point analysis previously defined; and (3) with the second residual to produce the program monitored by the third residual. This is a one-time pre-deployment cost, and it was found to take just a few seconds.
%that are randomised in a controlled way (if the program is run with $n$, and then with $m > n$ users, the user $i < n$ behaves the same in both runs, although possibly transactions may involve some of the new users, thus the time taken by the program increases with each new user.

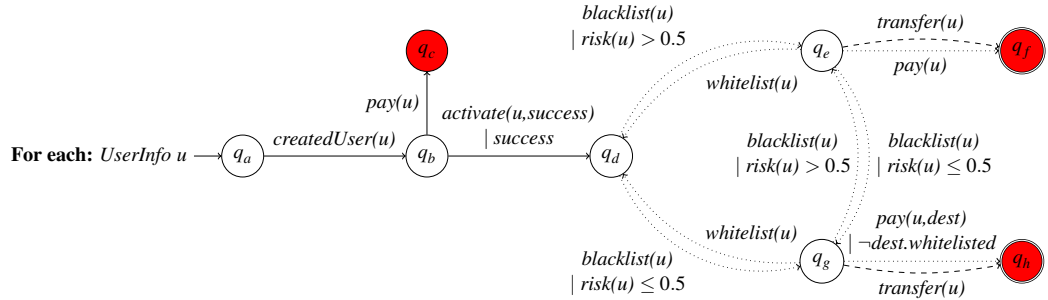
\begin{figure}[ptb]
\scalebox{0.7}{
  \begin{tikzpicture}[every text node part/.style={align=center},node distance=6cm,on grid,auto]
	%States
	%start
	\node (start) [initial, initial text={\textbf{For each:} \textit{UserInfo u}}, circle, draw] [] {$\q_a$};
%		\node (s) [] [above = 2cm of start] {\textbf{For each:} \textit{UserInfo u}};
	%loggedIn
	\node (created) [circle, draw] [right = 3.5cm of start]{$\q_b$};
	\node (createdBad) [circle,fill=red, draw] [above = 2cm of created]{$\q_c$};
	\node (activated) [circle, draw] [right = 3.5cm of created]{$\q_d$};
	%blacklistedlowrisk
	\node (inv1) [ circle] [above =2cm of activated]{};
	\node (blacklistedHighRisk) [circle, draw] [right = 4cm of inv1]{$\q_e$};
	%bad
	\node (blacklistedHighRiskBad) [accepting, circle, draw, fill=red] [right = 3.8cm of blacklistedHighRisk]{$\q_f$};
	%blacklistedlowrisk
		\node (inv2) [ circle] [below =2cm of activated]{};
	\node (blacklistedLowRisk) [circle, draw] [right = 4cm of inv2]{$\q_g$};
	%bad
	\node (blacklistedLowRiskBad) [accepting, circle, draw, fill=red] [right = 3.8cm of blacklistedLowRisk]{$\q_h$};
	
%	\node (ss) []  [left = 4cm of whitelisted]{};
%	\node (s_1) [circle, draw]  [below = 1cm of ss]{$\q_1$};
%	\node (s_1bad) [accepting, fill=red,circle, draw] [right = 3cm of s_1]{$\q_9$};

	Transitions
	\path[->] [] (start) edge node {\textit{createdUser(u)}} (created);
	\path[->] [] (created) edge node {\textit{activate(u,success)}\\$\mid \textit{success}$} (activated);
	\path[->] [] (created) edge node {\textit{pay(u)}} (createdBad);
	\path[->] [dotted] [bend left = 40] (activated) edge node {\textit{blacklist(u)}\\ $\mid$ \textit{risk(u)} $> 0.5$} (blacklistedHighRisk);
	\path[->,dotted] [left, bend right = 40] (activated) edge node {\\\\\textit{blacklist(u)}\\ $\mid$ \textit{risk(u)} $\leq 0.5$} (blacklistedLowRisk);
	\path[->,dotted] [below] (blacklistedHighRisk) edge node {\textit{pay(u)}} (blacklistedHighRiskBad);
	\path[->,dashed] [bend left = 10] (blacklistedHighRisk) edge node {\textit{transfer(u)}} (blacklistedHighRiskBad);
	\path[->,dotted] [bend right = 30] (blacklistedHighRisk) edge node {\textit{whitelist(u)}} (activated);
	\path[->,dotted] [right, bend left = 30] (blacklistedLowRisk) edge node {\textit{whitelist(u)}\\ \ } (activated);

	\path[->,dotted] [] (blacklistedLowRisk) edge node {\textit{pay(u,dest)}\\ $\mid \neg\textit{dest.whitelisted}$ } (blacklistedLowRiskBad);
		\path[->,dashed] [below, bend right = 10] (blacklistedLowRisk) edge node {\textit{transfer(u)}} (blacklistedLowRiskBad);
	\path[->] [dotted, bend right = 30] (blacklistedLowRisk) edge node {\textit{blacklist(u)}\\ $\mid$ \textit{risk(u)} $> 0.5$} (blacklistedHighRisk);
	\path[->] [dotted, bend left = 40] (blacklistedHighRisk) edge node {\textit{blacklist(u)}\\ $\mid$ \textit{risk(u)} $\leq 0.5$} (blacklistedLowRisk);
	\end{tikzpicture}  }
	\captionof{figure}{Property, with dashed transitions removed by the first analysis, and dotted by the third.%and dotted lines representing transitions removed by the first and third analysis respectively.
	}
  \label{fig:fitsproperty}
\end{figure}

The system was verified with respect to a specification constraining payment patterns based on the status of the user, e.g.\ a blacklisted user can only perform a payment if it has not exceeded a certain risk threshold. The risk level of a user is calculated by the monitor by checking that the companies the users deal with in general have transacted with users in good standing (i.e.\ that are not currently blacklisted or greylisted), with the number of such users in bad standing having an effect on the risk level of the user in question\footnote{Note that caching such a calculation does not aid the monitor performance since the risk level changes with each transaction --- also those not involving the user in question.}. This part of the specification is shown in \ref{fig:fitsproperty}, while \ref{fig:cfgs} illustrates the property-relevant behaviour of each possible user object, representing the system under analysis. %\gp{Is the time in ms or s? Fix axis appropriately}

%Different users exhibit different user-side behaviour, resulting in different CFGs depending on the status of the user in question. \ref{fig:cfgs} shows the abstracted behaviour for three kinds of users. 

%Note how we annotate the transitions in \ref{fig:fitsproperty} according to whether they would be removed given each of our analyses, while bronze users will not be monitored after the second analysis (since they cannot perform any payments, and thus cannot reach a bad state in the monitor).\cc{in the short version i would leave this out since we already say it in the caption.}

\begin{figure}[ptb]
\begin{minipage}{0.6\textwidth}
%\begin{figure}
%\scalebox{0.6}{
%\begin{tikzpicture}
%\begin{axis}[
%	title={},
%    xlabel={No. of Users},
%    ylabel={Percentage Overheads},
%    xmin=1000, xmax=1300,
%    ymin=0, ymax=200,
%    xtick={1000,1050,1100,1150,1200,1250, 1300},
%    ytick={0,20,40,60,80,100,120},
%    legend pos=north west,
%    ymajorgrids=true,
%    grid style=dashed,
%]
% %unmonitored
%\addplot[
%    color=blue,
%    mark=square,
%    ]
%    coordinates {
%    (1000,0)(1050,0)(1100,0)(1150,0)(1200,0)(1250,0)(1300,0)
%    };
% %monitored
%\addplot[
%    color=red,
%    mark=star,
%    ]
%    coordinates {
%    (1000,80.29)(1050,97.03)(1100,87.38)(1150,113.12)(1200,112.75)(1250,86.09)(1300,102.92)
%    };
%%QC
%\addplot[
%    color=green,
%    mark=triangle,
%    ]
%    coordinates {
%    (1000,79.43)(1050,92.46)(1100,88.29)(1150,114.09)(1200,109.86)(1250,79.74)(1300,107.73)
%    };
%%AQC
%\addplot[
%    color=violet,
%    mark=diamond,
%    ]
%    coordinates {
%    (1000,43.35)(1050,47.76)(1100,30.95)(1150,57.46)(1200,41.77)(1250,33.38)(1300,39.04)
%    };
%%ACFA
%{
%\addplot[
%    color=orange,
%    mark=otimes,
%    ]
%    coordinates {
%    (1000,9.2)(1050,1.69)(1100,4.51)(1150,2.35)(1200,4.68)(1250,2.19)(1300,4.32)
%    };
%}
%    \legend{Unmonitored,Monitored,1st Analysis,2nd Analysis,3rd Analsyis}
% 
%\end{axis}
%\end{tikzpicture}}
\scalebox{0.7}{
\begin{tabular}{|c|c|c|c|c|c|}
\hline
\begin{tabular}{@{}c@{}}\textbf{No. of}\\\textbf{Users}\end{tabular} & \textbf{Unmonitored} & \textbf{Monitored} & \begin{tabular}{@{}c@{}}\textbf{After}\\\textbf{1st}\end{tabular} & \begin{tabular}{@{}c@{}}\textbf{After}\\\textbf{2nd}\end{tabular} & \begin{tabular}{@{}c@{}}\textbf{After}\\\textbf{3rd}\end{tabular}
\\\hline 1000 & 206s & 371s & 369s & 295s & 225s
\\\hline 1050 & 231s & 456s & 445s & 342s & 235s
\\\hline 1100 & 24s & 450s & 452s & 314s & 251s
\\\hline 1150 & 254s & 542s & 544& 400s & 260s
\\\hline 1200 & 268s & 570s & 562s & 380s & 280s
\\\hline 1250 & 309s & 576s & 556s & 413s & 316s
\\\hline 1300 & 316s & 642s & 657s & 440s & 330s
\\\hline \hline 
\begin{tabular}{@{}c@{}}\textbf{Average}\\\textbf{Overheads}\end{tabular} &0 \%& 97.08\%	&95.94\%	& 41.96\% &	4.13\%\\\hline
\end{tabular}
}
%\captionof{figure}{Experiment results in time taken, and percentage overheads (lower and upper bounds).}
\label{table:eval}            
%\caption{Overheads(\%) for program before, and after monitoring with residuals}
%\end{figure}
\end{minipage}%
\begin{minipage}{0.5\textwidth}
%\begin{figure}
\scalebox{0.6}{
\begin{tikzpicture}
\begin{axis}[
    title={},
    xlabel={Number of users},
    ylabel={Execution time / s},
    xmin=1000, xmax=1300,
    ymin=100, ymax=900,
    xtick={1000,1050,1100,1150,1200,1250, 1300},
    ytick={0,200,300,400,500,600,700,900},
    legend pos=north west,
    ymajorgrids=true,
    grid style=dashed,
]
 %unmonitored
\addplot[
    color=blue,
    mark=square,
    ]
    coordinates {
    (1000,205.708)(1050,231.24633)(1100,240.024)(1150,254.29667)(1200,267.82833)(1250,309.23667)(1300,316.364)
    };
 %monitored
\addplot[
    color=red,
    mark=star,
    ]
    coordinates {
    (1000,370.877)(1050,455.632)(1100,449.768)(1150,541.967)(1200,569.80533)(1250,575.45933)(1300,641.95933)
    };
%QC
\addplot[
    color=green,
    mark=triangle,
    ]
    coordinates {
    (1000,369.101)(1050,445.065)(1100,451.93867)(1150,544.43233)(1200,562.07267)(1250,555.82167)(1300,657.17267)
    };
%AQC
\addplot[
    color=violet,
    mark=diamond,
    ]
    coordinates {
    (1000,294.889)(1050,341.688)(1100,314.30167)(1150,400.41233)(1200,379.69233)(1250,412.46567)(1300,439.87067)
    };
%ACFA
{
\addplot[
    color=orange,
    mark=otimes,
    ]
    coordinates {
    (1000,224.63267)(1050,235.15067)(1100,250.858)(1150,250.28167)(1200,280.359)(1250,316.01533)(1300,330.02233)
    };
}
    \legend{Unmonitored,Monitored,First analysis,Second analysis,Third analysis}
 
\end{axis}
\end{tikzpicture}
}
%\caption{Execution time for program before and after monitoring with residuals.}
%\end{figure}
\end{minipage}
\caption{Table and plot of the experiment results.}
\label{fig:results}
\end{figure}
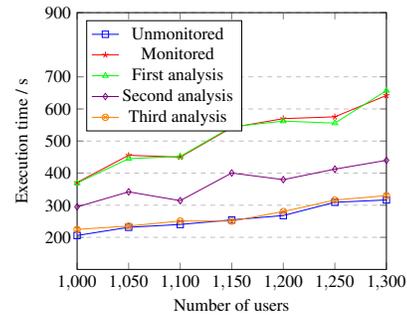

The memory used and execution time of the unmonitored program was compared to that of the monitored one, and to that of optimised monitors with the three analyses applied cumulatively. The experiment was run for different numbers of users, with three sample executions used to normalise differences between measurements. %ensure that the measurements are correct.%\gp{I have left out the users' CFGs, since we do not have the space for it here.}
Given that the monitors used in the experiment do not have state to keep track of, no measurable memory overheads were found any of the experiment runs. 
%\cc{the second graph is not really giving much more above the first... if I were a reviewer I would wish to see the graph continue to the right.}\sa{so which do you suggest we keep? I think the time taken graph shows clearly the pattern emerging and  I don't think a reviewer would wonder about the behaviour with different values. With the percentage graphs there could be more questions perhaps.} 
However, as expected, monitoring induced considerable processing overheads (see \ref{fig:results}), which were substantially reduced using our optimisations from an average of 97\% to just 4\%.%\gp{Add values here. We also need a table with the actual time and \% overhead.} 
 %and in terms of the percentage overheads at each stage in \ref{grp:percentages}. 

The first analysis took roughly the same time as the monitored time, since it only removed transitions that could never be taken, resulting in the monitoring engine only bypassing a conditional check for the never-activated event (since \textit{transfer} does not appear in \ref{fig:cfgs}). The second analysis did not reduce the property itself, but turned off monitoring of all statements in the program associated only with bronze users, avoiding monitors being created at runtime which are never violated (consider that in \ref{fig:cfgs} bronze users cannot effect payments). The third analysis identified that blacklisted users are never allowed to affect a payment by the application (i.e.\ the program never makes it to $q_f$ and $q_h$ in \ref{fig:fitsproperty}, consider the parallel composition of the CFGs and the property) and thus simply made sure that non-activated users do not affect payments (i.e.\ once a user was created, the monitor transitioned to $q_b$ and checked every incoming event against the only remaining outgoing transitions to $q_c$ and $q_d$). This resulted in an insignificant level of overheads, given the monitoring engine only had to check against two transitions (while in state $\q_c$), without any expensive conditions to check and no tight-looping. %\sa{I modified this paragraph slightly w.r.t \ref{fig:cfgs}}

Similar to the results from \cite{bodden}, the gains arise since the system does not necessarily use all the events appearing in the property and some of the correctness logic is encoded directly in the control-flow of the system. In our client-provider scenario: (i) the client does not make use of all the functions the provider allows (at least not for every possible object);
%\cc{object refer to the client, or objects the client uses?}
and (ii) the client is coded in such a way that allows reasoning about its control-flow e.g. blacklisting a user directly by setting a flag. In practice, we envisage that this approach is applicable, for instance, when encoding properties over APIs or constraining server-access, allowing for monitoring overhead reduction for API clients or clients accessing the server.

\section{Related Work}
\label{s:related-work}
%Other work exists that tries to reduce the overheads of runtime monitoring using static analysis. 
Our work builds directly on the results of Bodden et\ al.\ \cite{bodden}, but there are many other instances of the use of static analysis in order to optimise dynamic analysis. In \cite{isolares}, we previously presented a high-level theory of residuals, and a model-based approach to combining static and dynamic analysis, and gave informal examples of how residuals of DATEs could be computed. In this paper we present formally this intuition.

Dwyer et\ al.\ \cite{dwyer} take a different approach from ours or Clara's, wherein they identify safe regions in a program, i.e. sequences of statements that cannot violate a property, and if they are deterministic with respect to a property (if the monitor enters the region at a state $\q$ then it always exists at the same state $\q'$). The effect of the region on the monitor is then replaced by a new unique event $\e$, and the property augmented by a transition from $\q$ to $\q'$ with $\e$. Note, that this summarises the effect of some instrumentation into one, wherein we simply remove instrumentation that does not affect violation. %Note how this creates a property that is not a strict residual by our definition (since the property is being augmented by an event and transition not in the original property), but nonetheless does not effect violation of the property.
% The notion of identifying safe parts of a program also exists in model checking \cite{Lal2007}, which allows other composed analysis to focus simply on the parts that may contain an error, reducing the effort necessary to analyse the program.
Jin et\ al.\ \cite{parametricprops} also investigate parametric properties, and investigate optimisations which can be made to the implementation of monitoring logics at runtime, namely more efficient garbage collection of monitors associated with an object that has been garbage collected. It is worth noting how our second analysis may prevent some of this behaviour by detecting statically that an object may never violate and instead prevent the creation of its monitor. 
Other approaches try to make runtime overheads more predictable and manageable, but lose certainty of the verdict. \cite{Stoller2012} is an example of \emph{event sampling}, where not all events generated by a program are processed by the program, where this approach uses a statistical model to approximate the gaps in the execution trace. \cite{Bartocci2013} extends this approach to reduce the memory and time overheads incurred by statistical calculations performed at runtime, by estimating them statically.

%-\cite{raclet} define a quotient operator for component specifications, while they also use modal automata, which are automata with may and must functions from states to sets of actions, that specify actions that may or must occur at those states. This is different from our must and may functions that specify whether events program events can occur on the same object or not. 
\section{Conclusions and Future Work}
\label{s:conclusions}

Through this work, we have extended a static optimisation for the monitoring of parametric properties, to deal with automata with symbolic state. We reduce both the property using the system, and the system instrumentation by using the property. This static analysis is based on the control-flow of the program, and an aliasing relationship between relevant objects of the program, which we illustrated through a case study emulating a payment transaction system. These residual analyses have been implemented in a tool which we are currently preparing to make publicly available, this tool can be used to pre-process DATEs before they are used to monitor a program. We are in the process of combining these results with those of StaRVOOrS \cite{staticrv2}, where in contrast to our approach,  StaRVOOrS, static analysis is used to reduce the data-flow aspect of the specification (using pre- and post-conditions), leaving the control-flow aspect (in the form of DATEs) for dynamic analysis. Our work is complementary to this approach, and in fact we are currently investigating how to optimise properties using both control and data flow static analysis.% Future work also includes implementing the side-effect analysis we assumed in our work which would also involve the possibility of multiple DATEs monitoring the same program, at the same time, which brings into play consideration of the actions of the transitions of DATE, and how these effect the conditions of other transitions.

\nocite{*}
\bibliographystyle{eptcs}
\bibliography{references}
\end{document}